\documentclass[final]{elsarticle}
\usepackage{booktabs}
\usepackage{lineno,hyperref}
\usepackage{geometry}
\usepackage{multirow}
\makeatletter
\def\ps@pprintTitle{%
 \let\@oddhead\@empty
 \let\@evenhead\@empty
 \def\@oddfoot{}%
 \let\@evenfoot\@oddfoot}
\makeatother
\makeatletter
\renewcommand{\MaketitleBox}{%
  \resetTitleCounters
  \def\baselinestretch{1}%
  \begin{center}
    \def\baselinestretch{1}%
    \Large \@title \par
    \vskip 18pt
    \normalsize\elsauthors \par
    \vskip 10pt
    \footnotesize \itshape \elsaddress \par
  \end{center}
  \vskip 12pt
}
\makeatother

\usepackage{amsmath}
\usepackage{setspace}
\usepackage{subcaption}

\usepackage{nomencl}

\makenomenclature

\usepackage{ifthen}
  \renewcommand{\nomgroup}[1]{%
  \item[\bfseries
  \ifthenelse{\equal{#1}{N}}{Normal Symbols}{%
  \ifthenelse{\equal{#1}{A}}{Greek Symbols}{%
  \ifthenelse{\equal{#1}{C}}{Abbreviations}{%
  \ifthenelse{\equal{#1}{B}}{Subscripts}{%
  }}}}%
  ]}

\begin{document}
\doublespacing
\begin{frontmatter}
\title{\textbf{Radiative Heat Transfer Calculations using Full Spectrum k-Distribution Method for Benchmark Test Cases }}
\author{Kamal Khemani, Shreesh Parvatikar} 
\author{Pradeep Kumar}
\address{Numerical Experiment Laboratory (Radiation \& Fluid Flow Physics)\\
Indian Institute of Technology Mandi, Mandi, Himachal Pradesh, 175075, India}
\ead[]{pradeepkumar@iitmandi.ac.in}
\end{frontmatter}



\section*{Abstract}
In the present work, the full spectrum $k$-distribution method (FSK) has been adopted to calculate the radiative heat transfer in the presence of participating gaseous medium within an enclosure. The spectral radiative properties of the gaseous medium is obtained from the HITEMP-2010 database. Further, radiative properties have been assembled into a monotonically increasing function using the full spectrum $k$-distribution method. Moreover, a look-up table has been developed for these properties for different thermodynamic states of gases and a multi-dimensional linear interpolation technique for unavailable thermodynamic states of gases. Furthermore, the FSK method is extended for mixture of gases using different mixing models such as superposition, multiplication and  hybrid mixing model. The multiplication mixing model produces most accurate results among the mixture models used here. The results obtained from FSK has been validated against line by line method (LBL). The radiation transfer equation (RTE) is solved by finite volume method to calculate the wall heat fluxes and the divergence of radiative heat flux for various test cases in different category of homogeneous isothermal and isobaric and non-homogeneous non-isothermal non-isobaric media having different conditions of temperature pressure and mole-fraction. The FSK method has been successfully applied to non-homogeneous non-isothermal non-isobaric gaseous media for single gas or mixture of gases with almost LBL accuracy at extremely less computational cost and resource.
\section*{Keywords}
\noindent Non-gray radiation, line by line method, FSK look-up table, multidimensional linear interpolation, gas mixing models
 
\mbox{}

\nomenclature[]{$I_\eta$}{Spectral intensity}
\nomenclature[]{$s$}{Direction vector}
\nomenclature[]{$I_{b\eta}$}{Planck function}
\nomenclature[A]{$\kappa_\eta$}{Spectral absorption coefficient}
\nomenclature[A]{$\eta$}{Wavenumber}
\nomenclature[A]{$\sigma_{s\eta}$}{Spectral scattering coefficient}
\nomenclature[A]{$\beta_\eta$}{Spectral extinction coefficient}
\nomenclature[A]{$\Phi_\eta$}{Spectral scattering phase function}
\nomenclature[A]{$\epsilon_{w\eta}$}{Spectral emissivity of the wall}
\nomenclature[A]{$\Omega$}{Solid angle}
\nomenclature[]{$q$}{Radiative heat flux}
\nomenclature[A]{$\nabla \cdot q$}{Divergence of radiative heat flux}
\nomenclature[]{$G$}{Irradiation}
\nomenclature[]{$S_{if}$}{Line intensity}
\nomenclature[]{$k_{if}$}{Absorption crossection}
\nomenclature[A]{$\nu_{if}$}{Vacuum wavenumber}
\nomenclature[]{$T_{ref}$}{Reference temperature}
\nomenclature[]{$Q$}{Partition function}
\nomenclature[A]{$\gamma$}{Half width at half maximum}
\nomenclature[A]{$\gamma_{self}$}{Self broadened half width}
\nomenclature[A]{$\gamma_{air}$}{Air broadened half width}
\nomenclature[]{$p_{self}$}{Self pressure of gas}
\nomenclature[A]{$\delta$}{Line shift}
\nomenclature[]{$N$}{Number density of species}
\nomenclature[]{$f$}{Fractional Planck function}
\nomenclature[]{$k$}{re-ordered absorption coefficient}
\nomenclature[]{$g$}{cumulative $k$-distribution}
\nomenclature[]{$a$}{Stretching function}
\nomenclature[]{$w$}{weight for quadrature integration}
\nomenclature[]{$N_g$}{Number of participating gases}
\nomenclature[]{$p$}{Number of quadrature points}
\nomenclature[B]{$w$}{wall}
\nomenclature[B]{$\eta$}{wavenumber}
\nomenclature[B]{$ref$}{reference}
\nomenclature[B]{$g$}{gas}
\nomenclature[B]{$if$}{transition from upper state $f$ to lower state $i$}
\nomenclature[C]{HITEMP}{High temperature spectroscopic absorption parameter}
\nomenclature[C]{HITRAN}{High resolution transmission spectroscopic molecular absorption database}
\nomenclature[C]{CDSD}{Carbon-dioxide spectroscopic database}
\nomenclature[C]{LBL}{Line-by-line}
\nomenclature[C]{RTE}{Radiation transfer equation}
\nomenclature[C]{FVM}{Finite volume method}
\nomenclature[C]{WSGG}{Weighted sum-of-gray gas model}
\nomenclature[C]{DTM}{Discrete transfer model}
\nomenclature[C]{DOM}{Discrete ordinate method}
\nomenclature[C]{FSK}{Full spectrum $k$-distribution method}
\nomenclature[C]{ADF}{Absorption distribution function}
\nomenclature[C]{$P_N$}{Spherical harmonics method}
\nomenclature[C]{HWHM}{Half width at half maximum}
\nomenclature[C]{SMM}{Superposition mixing model}
\nomenclature[C]{MMM}{Multiplication mixing model}
\nomenclature[C]{HMM}{Hybrid mixing model}
\nomenclature[C]{SAM}{Spectral addition method}
\nomenclature[C]{SLW}{Spectral line weighted sum-of-gray gases model}
\nomenclature[B]{air}{air}
\nomenclature[B]{self}{self}

\printnomenclature

\section{Introduction}
The heat transfer by radiation mode plays a major role in many engineering applications, like, combustion, rocket propulsion, stealth vehicles, ablation of re-entry vehicles etc. Negligence or inaccurate treatment of radiation may lead to significant error in the estimation of the temperature profile, species distribution, pollutant emission or thermal protection. The heat transfer due to conduction and convection can be well determined theoretically or experimentally with a reasonable amount of accuracy and economy even for complex geometries. However, the same does not imply for radiation, as calculation of radiative heat flux even for simple two-dimensional geometries can be extremely difficult due to presence of participating media such as gases and particulate matters.
 
Gray medium is an engineering approximation where a constant value of absorption coefficient is assumed and the spectral variation of radiative properties are neglected. It simplifies the radiative transfer calculation, however, for some engineering applications like, combustion, rocket propulsion, it may lead to a significant amount of inaccuracy in predicting various quantities like flux, temperature distribution, species production etc. 
 
Unlike solids and liquids, gases do not emit and absorb energy in a continuous manner but, they emit and absorb energies in discrete bands at resonant frequencies which give dark or bright lines on a spectrum. Further, the spectral lines get broadened and partially overlap each other due to internal collisions and the other effects. This makes spectral absorption coefficient calculation a challenging task which in turn makes the calculation of radiation heat transfer extremely difficult. The accurate calculation of absorption coefficient can be performed using some of the most popular available databases like high resolution transmission spectroscopic molecular absorption database (HITRAN) \cite{rothman2009hitran}, carbon-dioxide spectroscopic database (CDSD) \cite{tashkun2011cdsd}, high temperature spectroscopic absorption parameter (HITEMP) \cite{rothman2010hitemp,hargreaves2019spectroscopic} etc.

Several non-gray models evolved over time and can be classified into three different categories namely (1) Line by line, (2) Band models, (3) Global models. Line-by-line method (LBL) is the most accurate non-gray method, as it solves the RTE for each spectral absorption coefficient. Even after advancements in computers, LBL calculations are used only for benchmarking purposes. As radiation heat transfer is a small part of big multi-physics problems, like combustion where it is required to solve for mass and species conservation for species transport, reaction chemistry, fluid flow, turbulence and their interactions with other modes of heat transfer coupled together, thus special consideration is needed for radiation heat transfer to reduce it's resource requirement.

The next class of non-gray model which gained a large amount of popularity is the band model. This model does not solve RTE for each spectral line, instead splits the whole spectral scale into a number of bands. Further, development in the field of band models leads to the development of narrow band and wide band models.

The narrow band model was first developed by Malkmus and Goody \cite{malkmus1967random,goody1952statistical}, where the actual variation of absorption coefficient is smoothened by taking an average over the narrow band. It gives good accuracy and is cheaper than LBL but it cannot be applied to non-homogeneous medium which is the biggest limitation of this method. The wide band model introduced by Edward \cite{edwards1976molecular} requires averaging to be done over the whole ro-vibrational band and its accuracy is even lesser than narrow band model. It considers this fact that the black body intensity does not vary significantly over the whole spectral range, thus, its correlations can be developed by integrating the narrow band over the entire spectrum.

The global models are one of the best available non-gray models in terms of accuracy and computational resource requirement, some of the most commonly used global models are weighted sum of gray gas model (WSGG), spectral line weighted sum-of-gray gases model (SLW), absorption distribution function model (ADF), full Spectrum k-distribution method (FSK). The weighted sum of gray gas model was initially developed by Hottel and Sarofim \cite{hottelradiative}, as the name suggests, it relates the non-gray gas into number of gray gases having different weights and thus, solves RTE for different gray gases. Nevertheless, this method is restricted to black wall enclosures containing non-scattering medium. This model is further improved by spectral line weighted sum-of-gray gases model including non-isothermal and non-homogeneous medium by introducing cumulative distribution of absorption coefficient into weighted fractional Planck function over the entire spectrum by Dension and Webb \cite{denison1995spectral}. The absorption distribution function model (ADF) \cite{pierrot1999fictitious} differs from SLW only in terms of weight considered for the gray gases. The full Spectrum k-distribution method which can be used extensively for non-homogeneous medium by correlating or scaling the absorption coefficient, developed by Modest and Zhang \cite{modest2002full} is one of the best available non-gray radiation model and can be used for scattering media with gray walls. This method is further improved by considering the conservation of emission \cite{cai2014improved}. It is also stated by Modest that WSGG is only a crude implementation of FSK. However, co-relating or scaling the absorption coefficient on-the fly requires tremendous amount of computational resource. Due to these problems, Wang et. al \cite{wang2016full} developed a FSK database which contains pre-calculated value of re-ordered absorption coefficient at different thermodynamic states of gases and soot particle. They further improved the accuracy of database \cite{wang2019full} by incorporating a new scheme to store correlated values, they further compared the accuracy of new database. The radiative properties at a particular thermodynamic state is retrieved using a multi-dimensional linear interpolation technique. A technique to calculate non-gray stretching factor on the fly was also developed, this drastically reduced the size of database. Zhou et al. \cite{zhou} used the previously developed FSK database by Wang et al. \cite{wang2019full} and developed a machine learning based algorithm to obtain properties without performing on the fly 6D interpolation. The size of database developed by Wang is quite large, i.e., 3.2 GB which makes data retrieval a complex and time taking task but the machine learning model developed by Zhou is only 32 MB. They tested the model for different scenarios and the results obtained matches well with the previous database with higher computational efficiency. Zheng et al. \cite{zheng2020improved} developed a new FSCK method namely FSCK-RSM (response surface mdoel). They improved the computational efficiency of conventional FSCK to extract k-distribution by using a radial basis function which helps to avoid multiple computations and interpolations. They further checked the computational resource requirements with different methods and concluded that FSCK-RSM needs 677 times less computational resource when compared with conventional FSCK. Liu et al. \cite{liu2022development} developed a look-up table for multigroup full spectrum k-distribution (MG-FSK) for a large number of thermodynamic states of $CO_2$ and $H_2O$. MG-FSK provides higher accuracy then conventional FSCK look up table for non-isothermal gaseous medium. Author's tested this look-up table for various test cases and concluded that it is highly accurate and the relative error is not more than 1\%. Global models can be used easily with any of the available RTE solvers without any modifications. Some of the most popular numerical techniques to solve RTE are spherical harmonics $P_N$ method, discrete ordinate method (DOM), discrete transfer method (DTM), finite volume method (FVM), Monte Carlo method etc.

In the present study, the calculation of non-gray radiative heat transfer has been performed for non-homogeneous and non-isothermal participating medium in an enclosure. The medium is filled with the mixture of participating gases such as $H_2O$ and $CO_2$ at different thermodynamic states of temperature and mole-fraction. The radiative properties are evaluated from HITEMP-2010 database and is reordered using full spectrum k-distribution method. Further, the FSK database (FSK look-up table) is generated to make it applicable for the non-homogeneous and non-isothermal medium by pre-calculating the properties and storing it for different thermodynamic states. An efficient scheme of interpolation has also been presented to evaluate the value of radiative properties which are not available in the look-up table. Several mixing models have also been presented to obtain $k$-distribution for mixture of participating gases for both homogeneous isothermal medium and non-homogeneous non-isothermal medium.\\
The manuscript is organised as section 2 describing the detailed mathematical models to obtain spectral absorption coefficient, full spectrum $k$-distribution for individual gas and gas mixtures, followed by results and discussions in section 3, and finally the present work is concluded in section 4.
\section{Mathematical Model}
\subsection{The spectral radiative transfer equation}
\noindent The spectral radiative transfer equation (s-RTE) for the participating medium can be obtained by making spectral radiative energy balance for a pencil ray that passes through the medium. A ray gets attenuated due to absorption and out-scattering and augmented by emission and in-scattering. Considering all the above phenomena, the equation of radiation transfer for a monochromatic ray is given as \cite{modest2013radiative,howell2020thermal},
\begin{equation} \label{RTE}
\
\frac{dI_{\eta}}{ds}=\kappa_{\eta}I_{b\eta}-\beta_{\eta}I_{\eta}+\frac{\sigma_{s\eta}}{4\pi}\int_{4\pi}I_{\eta}(\hat{s_{i}})\:\Phi_{\eta}(\hat{s_{i}},\hat{s})\:d\Omega_{i} \:,
\
\end{equation}
Above equation is subjected to wall boundary condition,
\begin{equation} \label{bcrte}
    I_\eta(r_w,\hat{s})=\epsilon_{w\eta}I_b(r_w)+\frac{1-\epsilon_{w\eta}}{\pi}\int_{\hat{n}\cdot\hat{s}>0}I_\eta(r_w,\hat{s})\:|\hat{n}\cdot\hat{s}|\: d\Omega \quad for \quad (\hat{n}\cdot \hat{s}<0)
\end{equation}
where, $\epsilon_{w\eta}$ is the spectral wall emissivity, $I_\eta$ is the spectral intensity along $\hat{s_{i}}$, $I_{b\eta}$ is the Planck function, $\kappa_{\eta}$ is the spectral absorption coefficient, $\eta$ is the wavenumber, $\sigma_{s\eta}$ is the spectral scattering coefficient, $\beta_{\eta}$ is the spectral extinction coefficient, $\Phi_{\eta}$ is the spectral scattering phase function which gives the probability of ray getting scattered from one direction to other and $\Omega$ is the solid angle.\\
Although above (Eq. \ref{RTE}) considers all the phenomena of radiation, however, only absorption and emission phenomena is a gaseous medium are considered in the present work. Thus, the general RTE for above two phenomena is written as,

\begin{equation}\label{absoemitRTE}
\
\frac{dI_{\eta}}{ds}=\kappa_{\eta}I_{b\eta}-\kappa_{\eta}I_{\eta}
\
\end{equation}
  However, it is subjected to same boundary condition as Eq.(\ref{bcrte}).
  The solution of above equation will provide the spectral intensity field in different directions. The radiative flux can be calculated from the spectral intensity field as, 
\begin{equation}  \label{radflux}
    q =\int_0^\infty q_\eta \: d\eta=\int_0^\infty \int_{4\pi} I_\eta \:|\hat{n}\cdot\hat{s}|\:d\Omega\: d\eta
\end{equation}
The temperature field is obtained from the energy equation which considers the contribution of all the modes of heat transfer and radiation contribution is included as the divergence of radiative heat flux which is a sink term in the total energy equation.
Physically, the divergence of radiative heat flux represents the net loss of radiative energy from a control volume, which is equal to the difference of energy emitted and absorbed in the medium. Thus, divergence of radiative flux is given as,
\begin{equation} \label{divRTE}
    \nabla \cdot q = \int_0^\infty \kappa_\eta\left(4\pi I_{b\eta}-\int_{4\pi}I_\eta \: d\eta \right) d\eta =\int_0^\infty \kappa_\eta(4\pi I_{b\eta}-G_\eta)\: d\eta
\end{equation}
The s-RTE (Eq.\ref{RTE}) is an integro-differential equation having six independent variables which includes three spatial coordinates, two angular directions and a wavenumber making it extremely difficult to solve for complex geometries. The complexity of the problem increases many-fold with the participating gases in the medium, due to the fact that the radiative properties of gases like, absorption and scattering coefficient vary abruptly with the electromagnetic spectrum containing millions of spectral lines for a single thermodynamic state. A proper estimation of these properties are important for calculation of heat transfer by radiation mode accurately.\\
These properties can be obtained from the knowledge of spectroscopy, however, high temperature spectroscopic absorption parameter (HITEMP) \cite{rothman2010hitemp} is one of the most popular available database used to determine radiative properties, like absorption coefficient precisely. It is developed and maintained by the "Atomic and Molecular Physics Division, Harvard-Smithsonian Center for Astrophysics". It contains the spectroscopic information of species in 160 characters in ASCII format depicting different spectroscopic parameters. The short information about the species available in this database is given in Table 1. The information is stored at the reference temperature of 296 K and pressure 1 atm. A calculation is required to obtain the spectroscopic information at different thermodynamic state and this is accurate for temperature upto 4000 K and pressure 80 bar for species available in the database.

\begin{table}[]
\centering
\caption{Content of HITEMP-2010 database}
\begin{tabular}{@{} l *3c @{}}
\toprule
\textbf{Molecule} &
  \textbf{\begin{tabular}[c]{@{}c@{}}Spectral \\ Coverage ($cm^{-1}$)\end{tabular}} &
  \textbf{\begin{tabular}[c]{@{}c@{}}Number of\\ isotopologues\end{tabular}} &
  \textbf{\begin{tabular}[c]{@{}c@{}}Number of \\ transitions\end{tabular}} \\ \midrule
$H_2O$ & 0-30000 & 6 & 114,241,164 \\ 
$CO_2$ & 5-12785 & 7 & 11,193,608  \\ 
$CO$   & 0-8465  & 6 & 113,631     \\ 
$NO$   & 0-9274  & 3 & 115,610     \\ 
$OH$   & 0-19268 & 3 & 41,557      \\ \bottomrule
\end{tabular}
\end{table}


\subsection{Calculation of absorption coefficient from HITEMP-2010 database}

The spectral absorption coefficient of participating gases like $CO_2$, $H_2O$, $CO$, $NO$ and $OH$ are calculated from HITEMP-2010 \cite{rothman2010hitemp}. The LBL approach has been used to calculate the spectral absorption coefficient with a uniform spectral spacing of 0.01 cm$^{-1}$. The transition line intensity other than reference thermodynamic state of a species is obtained by following equation.

\begin{equation} \label{lineInt}
\
S_{if}(T)=S_{if}(T_{ref})\frac{Q(T_{ref})}{Q(T)}\frac{exp(-c_{2}E/T)}{exp(-c_{2}E/T_{ref})}\left[\frac{1-exp(-c_{2}\nu_{if}/T)}{1-exp(-c_{2}\nu_{if}/T_{ref})}\right]\:,
\
\end{equation}
where $c_2$ is second radiation constant, $T_{ref}$ is the reference temperature, i.e., 296 K, $S_{if}(T)$, $\nu_{if}$ is the vacuum wavenumber, $E$ is the lower state energy and  $Q(T_{ref})$/$Q(T)$ is the ratio of internal partition functions. These parameters are obtained from HITEMP database.\\
\noindent The collisional broadening phenomenon is used in the current study and this broadening phenomenon is applicable for temperature upto 2000 K and at or above atmospheric pressure. The absorption crossection obtained from line intensity and this broadening phenomenon is shown below,

\begin{equation}
    \
k_{if}(\nu,T,p)=\left(\frac{S_{if}(T)}{\pi}\right)\left[\frac{\gamma(p,T)}{\gamma^{2}(p,T)+(\nu-\nu_{if}^{*})^{2}}\right] \:,
\
\end{equation}
where, $k_{if}$ is the monochromatic absorption coefficient and $\gamma(p,T)$ is half width at half maximum (HWHM) at given temperature and pressure which can be calculated as 
\begin{equation}
\
\gamma(p,T)=\left(\frac{T_{ref}}{T}\right)^{n_{air}}(\gamma_{air}(p_{ref},T_{ref})(p-p_{ref})+\gamma_{self}(p_{ref},T_{ref})p_{self}) \:,
\
\end{equation}
\begin{equation}
\
\nu_{if}^{*}=\nu_{if}+\delta(p_{ref})p\:,
\
\end{equation}
where n$_{air}$ is temperature-dependence coefficient, $\gamma_{self}$ is self-broadened half width, $\gamma_{air}$ is air broadened half width, $p_{self}$ is the partial pressure of gas, $\delta$ is shift of line due to pressure.\\
\noindent Finally, the absorption coefficient is obtained as,

\begin{equation}
\
\kappa_{\eta}=k_{if}*N
\
\end{equation}
where $N$ is number density of species which is given as 
\begin{equation}
\
N=(7.3392315\times10^{21})\times p_{self}/T
\
\end{equation}
Algorithm for calculation of spectral absorption coefficient \cite{bartwal2017calculationCO2,bartwal2017calculationH2O}
\begin{itemize}
    \item[1.] Extract the required parameters from HITEMP-2010 database.
    \item[2.] Calculate line intensity at required temperature from Eq.(\ref{lineInt}).
    \item[3.] Evaluate Half width at half maxima (HWHM) using Lorentz profile.
    \item[4.] Calculate the shift of line due to pressure.
    \item[5.] Now create an equispaced division having spectral interval of 0.01 cm$^{-1}$ for whole spectrum.
    \item[6.] Consider the effect of transition upto a spectral distance of 20 HWHM from both the sides at a particular spectral location \cite{chu2014calculations}.
    \item[7.] Evaluate monochromatic absorption coefficient and calculate absorption coefficient by multiplying it with molecule number density ($N$) of the radiating species.
    \item[8.] Repeat steps 6 to 7 at all spectral location.
    \item[9.] Filter the data to remove the noise.
\end{itemize}

In a general engineering system, there hardly exists a single participating gas, e.g., in combustion of any hydrocarbon fuel, three major participating gases are formed, i.e., $CO_2$, $H_2O$ and $CO$. So, it requires calculation of spectral absorption coefficient for mixture of gases  \cite{bartwal2018calculation} and it can be obtained as, 

\begin{equation} \label{eq:mixAbso}
    \kappa_{\eta,mix}=\sum_{i=1}^m \kappa_{\eta i}
\end{equation}
where $m$ is the number of participating gases \\

At this point, one can imagine that there exists millions of transition lines and Eq. (\ref{absoemitRTE}) needs to be solved on these millions of lines in each direction for a single thermodynamic state of gas. Furthermore, there exists many thermodynamic state of gases in any engineering applications like, combustion, plume radiation, gasification etc. Thus, it is formidable to use this method in the current form for any applications with the current state of the art computer technology. Many methods have been proposed to reduce the computational resource requirement as much as possible. The more realistic and accurate methods are FSK \cite{modest2002full,cai2014improved} and SLW \cite{denison1995spectral} methods developed for single thermodynamic state and it is further improved for many thermodynamic states that exists in a system.
The FSK method for RTE calculation for a single thermodynamic state is described below.
\subsection{Full spectrum k-distribution method}
The spectral absorption coefficient obtained from HITEMP-2010 database is highly erratic in nature and attains same value multiple times. So, within a narrow band, where Planck function is essentially constant, it can be re-ordered into monotonically increasing function.
This reduces the computational effort to solve RTE over few quadrature points which provides same accuracy as solving RTE by LBL method. This method of re-ordering of spectral absorption coefficient is called the full spectrum $k$-distribution (FSK) method and is exact for homogeneous medium. The details about this method are given below. \\

\subsubsection{Homogeneous isothermal and isobaric medium}
\noindent 
The full spectrum $k$-distribution which accounts the fractional Planck function between two absorption coefficients, i.e., $\kappa_\eta$ and $\kappa_{\eta+\Delta \eta}$ as $\Delta \eta \rightarrow 0$ is given as,
\begin{equation}
\
f(T,k)=\frac{1}{I_{b}}\int_{0}^{\infty}I_{b\eta}\:\delta(k-\kappa_{\eta})d\eta
\
\end{equation}
where $f(T,k)$ is Planck function weighted $k$-distribution. The spectral absorption coefficient varies between zero to few hundreds over whole spectral range where lower values of absorption coefficients are large in number, thus, it is not advisable to have a uniform division of spectral absorption coefficient scale as it does not capture lower absorption coefficient values efficiently. Therefore, division of absorption coefficient should happen according to a power law distribution \cite{wang2004high}. So, absorption coefficient selection over a range given below as,

\begin{equation}   \label{powr}
    \
k=\left[k_{min}^{pow}+\frac{i-1}{N_{bin}-1}(k_{max}^{pow}-k_{min}^{pow})\right]^{1/pow}
\
\end{equation}
where, $k_{min}$ and $k_{max}$ are the minimum and maximum values of spectral absorption coefficients, respectively. Value of $pow$ is taken as 0.3678 and the spectral axis is divided into 5000 bins $(N_{bin})$. 
It is further advisable to develop cumulative $k$ distribution, i.e., $g$ which is a smooth increasing function whose value lies from 0 to 1 and is obtained as,
\begin{equation}
\
g(T,k)=\int_{0}^{k}f(T,k)\:dk \: ,
\
\end{equation}
Finally, the reordered radiative transfer equation (r-RTE) for absorbing-emitting medium in $g$ space is given as,
\begin{equation} \label{eq:Rte in g space}
\
\frac{dI_{g}}{ds}=k\:(I_{b}-I_{g}) \: ,
\
\end{equation}
subjected to wall boundary condition,
\begin{equation}
\
I_{w g}=\epsilon_w \: a(T_w,T,g)\:\sigma T_w^4 + \frac{(1-\epsilon_w)}{\pi} \int_{(\hat{n}\cdot\hat{s})<0}I_{g}\:|\hat{n}\cdot\hat{s}|\: d\Omega \quad for \quad (\hat{n}\cdot \hat{s}>0)
\
\end{equation}
The detailed mathematical derivation to obtain r-RTE from s-RTE can be referred from \cite{modest2013radiative,howell2020thermal}. The transformed boundary condition contains a stretching function $a(T_w,T,g)$ which can be calculated as,
\begin{equation}
\
a(T_w,T,g)=\frac{f(T_w,k)}{f(T,k)}=\frac{dg(T_w,k)}{dg(T,k)}
\
\end{equation}
It is the ratio of two $k$-distribution functions which are calculated at wall temperature and medium temperature, respectively. Further, it is the ratio of gradient of $g$ function at wall and medium temperature.
\noindent Finally, total intensity can be obtained as,
\begin{equation} \label{eq:intensity final}
    I=\int_0^\infty I_\eta \: d\eta=\int_0^\infty I_k \: dk=\int_0^1 I_g \: dg \: ,
\end{equation}
$g$ can also be termed as non-dimensional Planck weighted cumulative wavenumber and this transformation of RTE from $\eta$ space to $g$ space is called as full spectrum $k$-distribution method. Further,
total radiative heat flux and divergence of total radiative flux in $g$ space is given as,
\begin{equation}
    q=\int_0^1 q_g \: dg = \int_0^1 \int_{4\pi} I_g\:(\hat{s})\:|\hat{n} \cdot \hat{s}|\:d\Omega \: dg\: ,
\end{equation}
\begin{equation}
        \nabla \cdot q = \int_0^1 k_g(4\pi I_{b}-G_g)\: dg\: ,
\end{equation}


After constructing $k$-distribution function in $g$ space, it is required to select quadrature points on $k$-distribution function over which r-RTE needs to be solved. The strategy to discretize $k$-distribution function is to fix quadrature points for $g$ and corresponding $k$ values can be obtained. Similarly non-gray stretching factor is also obtained for the same number of quadrature points \cite{wang2004high,davis2007methods}.\\
The $k$-distribution function is not a regular function, thus, the selection of quadrature points on $g$ scale should be such that the most of the points lie on higher values of $k$. Therefore, quadrature points on $g$ scale can be selected as ,
\begin{equation}
    \
    g_n=cos(\theta_n)\: ,
    \
\end{equation}
where,
\begin{equation}
    \
    \theta_n=\left(\frac{n\pi}{2p+1}\right) \qquad n=1,2,..,p
    \
\end{equation}
where $p$ is the number of quadrature points which can be 12, 16, 32 or any. It is necessary to have more number of $k$ points at higher $g$ side as higher values of absorption coefficient dominates the radiation transfer. The corresponding weights $w$ for integral can be calculated as,
\begin{equation}
    \
    w_n=\frac{4\:sin(\theta_n)}{(2p+1)}\sum_{t=1}^{p}\frac{sin\:(2t-1)\:\theta_n}{(2t-1)}
    \
\end{equation}
Eq. (\ref{eq:intensity final}) for total intensity can now be written as 
\begin{equation}
\
I=\int_{0}^{1}I_g \: dg=\sum_{i=1}^{p}\; I_{gi}w_i
\
\end{equation}
Finally, algorithm to develop FSK distribution function as 
\begin{itemize}
     \item[1.] Calculate spectral absorption coefficient of a gas from HITEMP-2010 database.
    \item[2.] The spectral absorption $\kappa_\eta$ axis is divided into large number of bins, say 5000 between $k_{min}$ to $k_{max}$ Eq. (\ref{powr}). More number of bins leads to smoother $k$-distribution.
    \item[3.] Evaluate the fractional Planck function between two $k$ bins over whole spectrum.
    \item[4.] Further, add the fractional Planck function cumulatively to get $g$ distribution. 
    \item[5.] Calculate fixed quadrature points on $g$ scale and weight $w$ for quadrature points required for integration.
    \item[6.] Obtain $k$ and $a$ values from fixed quadrature points on $g$ scale.
\end{itemize}

\subsubsection{Non-homogeneous non-isothermal and non-isobaric medium}

The FSK method is exact for homogeneous medium. But, engineering problems are seldomly homogeneous. To apply above formulation of FSK for non-homogeneous non-isothermal and non-isobaric medium, where many thermodynamic state exists, correlation (Full spectrum correlated $k$-distribution) or scaling (Full spectrum scaled $k$-distribution) of absorption coefficient can be employed \cite{modest2002full,cai2014improved} or pre-calculate the reordered absorption coefficient and store in the form of a look-up table for different thermodynamic states of gases. We follow the later strategy \cite{wang2016full,wang2019full} to solve RTE for non-homogeneous and non-isothermal medium because FSCK may lead to significant error at higher temperature due to presence of hot lines.\\ 
Currently, the FSK database is generated for $H_2O$ and $CO_2$ at pressure of 1 bar, temperatures ranging from 300 to 2000 K and mole-fractions varying from 0 to 1. The database contains reordered absorption coefficient $k$ and non-gray stretching factor $a$ for 32 points. The information of data for different thermodynamic state is summarized in Table 2. \\
After generating look-up table, an efficient scheme for interpolation has been utilized to calculate the values of unavailable thermodynamic state. The inverse lever rule is used for single variable, i.e., either temperature or mole-fraction interpolation given below as,


\begin{equation}
    k(x)=\frac{1}{(x_{2}-x_{1})}(k(P_{1})(x_{2}-x)+k(P_{2})(x-x_{1}))
\end{equation}
The bi-linear interpolation method is used for the two variable interpolation i.e., temperature and mole-fraction as given below,
\begin{multline}
 k(x,t)=\frac{1}{(x_2-x_1)(t_2-t_1)} (k(P_{11})(x_{2}-x)(t_{2}-t)+k(P_{21})(x-x_{1})(t_{2}-t)+k(P_{12})(x_{2}-x)(t-t_{1})\\+k(P_{22})(x-x_{1})(t-t_{1}))   =\frac{1}{(x_{2}-x_{1})(t_{2}-t_{1})}[x_{2}-x \quad x-x_{1}]\begin{bmatrix}k(P_{11}) & k(P_{12})\\
k(P_{21}) & k(P_{22})
\end{bmatrix}\begin{bmatrix}t_{2}-t\\
t-t_{1}
\end{bmatrix}
\end{multline}
where, $P_{11}=(x_1,t_1),\  P_{12}=(x_1,t_2), \ P_{21}=(x_2,t_1)$ and $P_{22}=(x_2,t_2)$ are the four combinations of temperature and mole-fraction, i.e., two variable interpolation.

\begin{table}[]
\centering
\caption{Information of data stored in FSK lookup table for different thermodynamic states}
\begin{tabular}{@{} l *3l @{}}
\toprule
\textbf{Parameters} & \textbf{Range}   & \textbf{Interval} & \textbf{No. of data points} \\ \midrule
Pressure            & 1-10 bar          &  Every 1 bar  & 10                     \\ \\
Gas Temperature     & 300-2000 K       & Every 100 K     & 18                    \\ \\
\begin{tabular}[l]{@{}l@{}}Mole fraction of $CO_2$    \end{tabular}  &
  \begin{tabular}[l]{@{}l@{}}0-0.1\\ 0.25-1\end{tabular} &
  \begin{tabular}[l]{@{}l@{}}Every 0.05\\ Every 0.25\end{tabular} &
  7   \\ \\
\begin{tabular}[l]{@{}l@{}}Mole fraction of $H_2O$  \end{tabular} & 
  \begin{tabular}[l]{@{}l@{}}0-0.05\\ 0.1-0.2\\ 0.25-1\end{tabular} &
  \begin{tabular}[l]{@{}l@{}}Every 0.01\\ Every 0.05\\ Every 0.25\end{tabular} &
  13 \\ \bottomrule
\end{tabular}
\label{table:lookup}
\end{table}

\subsection{Full spectrum $k$-distribution method for mixture of gases}
Many engineering problems contains mixture of participating gases rather a single participating gas. To obtain the exact $k$-distribution for the mixture of gases, we need to calculate spectral absorption coefficient for mixture of gases Eq. (\ref{eq:mixAbso}), but this requires prior knowledge of all the thermodynamic states available in the problem. The calculation of mixture $k$-distribution from individual $k$-distribution produces a large error \cite{modest2005assembly} as the spectral line information is lost while assembling $k$-distribution from single gas $k$-distribution. Nevertheless, several mathematical modelling \cite{modest2005assembly,solovjov2000slw} have been developed to obtain mixture $k$-distribution from individual gas $k$-distribution, some of them are as following

\subsubsection{Superposition mixing model (SMM):}
It is based on the assumption, that the spectral lines do not overlap significantly, leading to addition of $g$ values for mixture of gases
\begin{equation}
    g_{mix}(k_{mix})=\sum_i^{N_g} g_i(k_i) - (N_g-1)
\end{equation}
where, $g_i$ is the $g$-distribution of individual gas and $N_g$ is the number of participating gases. $k_{mix}$ is the mixture $k$-distribution and can be obtained using power law Eq. (\ref{powr}) with $k_{min}$ as the sum of minimum value of spectral absorption coefficient and $k_{max}$ as the sum of maximum value of spectral absorption coefficient of gases.

\subsubsection{Multiplication mixing model (MMM):}
This model describes the probability of individual event, of uncorrelated absorption coefficients for the gases \cite{solovjov2000slw}. The mixture $g$-values can be obtained by multiplying $g$-values of individual gas.
\begin{equation}
    g_{mix}(k_{mix})=\prod_i^{N_g} g_i(k_i)
\end{equation}
$g_i$ is the $g$-distribution of individual gas and $N_g$ is the number of participating gases.

\subsubsection{Hybrid mixing model (HMM):}
The SMM gives good results at higher value of absorption coefficient while the MMM gives good results at lower value of absorption coefficient \cite{solovjov2000slw}, thus the convolution of above two is the hybrid model and is given as,
\begin{equation}
    g_{mix}(k_{mix})=\frac{[g_{sup}(k_{mix})\times log(k_{mix}/k_{max})+g_{mult}(k_{mix})\times log(k_{min}/k_{mix})]}{log(k_{max}/k_{min})}
\end{equation}
where, $g_{sup}(k_{mix})$ and $g_{mult}(k_{mix})$ is the mixture $g$ obtained from SMM and MMM respectively.

\section{Results and Discussions}
The above procedures have been employed to calculate the spectral absorption coefficients of individual $H_2O$ and $CO_2$ gas and mixture of these gases. Further, the results of radiative transfer equation with FSK for single thermodynamic state and many thermodynamic states have been present and computational time is reported in the following sections.
\subsection{Spectral absorption coefficient of $H_2O$ and $CO_2$ gases}

The spectral absorption coefficients of individual $H_2O$ and $CO_2$ gases have been calculated by LBL approach from HITEMP-2010 database and is depicted in Figure 1 and 2, respectively. The absorption of energy occurs in three distinct spectral bands for $H_2O$ which corresponds to 6.3$\mu$m, 2.7$\mu$m and 1.8$\mu$m bands. The 6.3$\mu$m band is called strong fundamental band, 2.7$\mu$m is first overtone band and 1.8$\mu$m is second overtone band. Similarly, $CO_2$ absorbs in four distinct bands, which corresponds to 2$\mu$m, 4.3$\mu$m, 2.7$\mu$m and 15$\mu$m. The 4.3$\mu$m band is the fundamental band. The $H_2O$ and $CO_2$ act as transparent gases for rest of the spectrum. The description of these bands for $H_2O$ and $CO_2$ gases is shown in Table 3. The spectral absorption coefficients for $H_2O$ and $CO_2$ on the log scale are shown in Figure 3(a) and 3(b), respectively.

\begin{figure}
    \centering
    \includegraphics[scale=0.35]{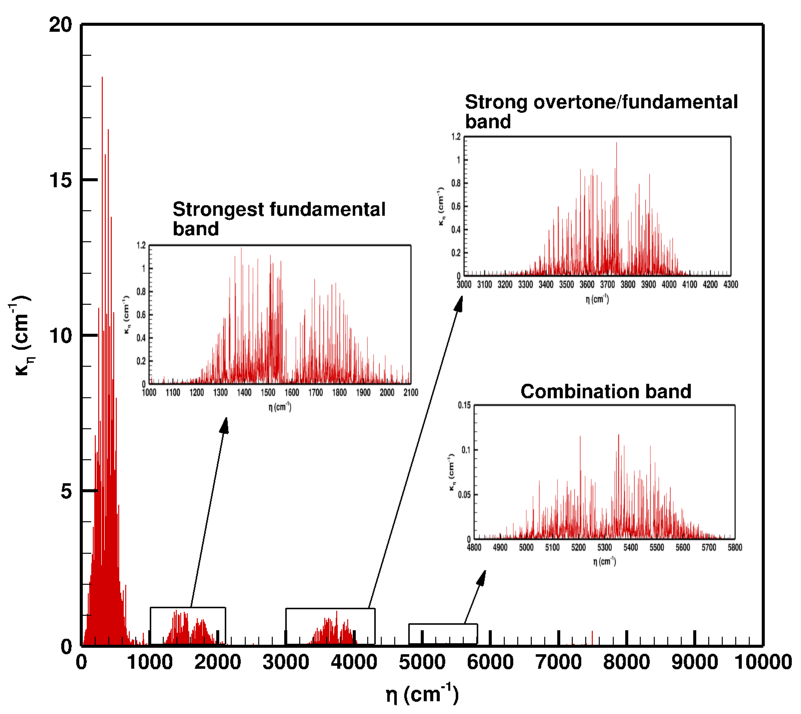}
    \caption{Spectral absorption coefficient of pure $H_2O$ at 1000 K and 1 atm}
\end{figure}

\begin{figure}
    \centering
    \includegraphics[scale=0.35]{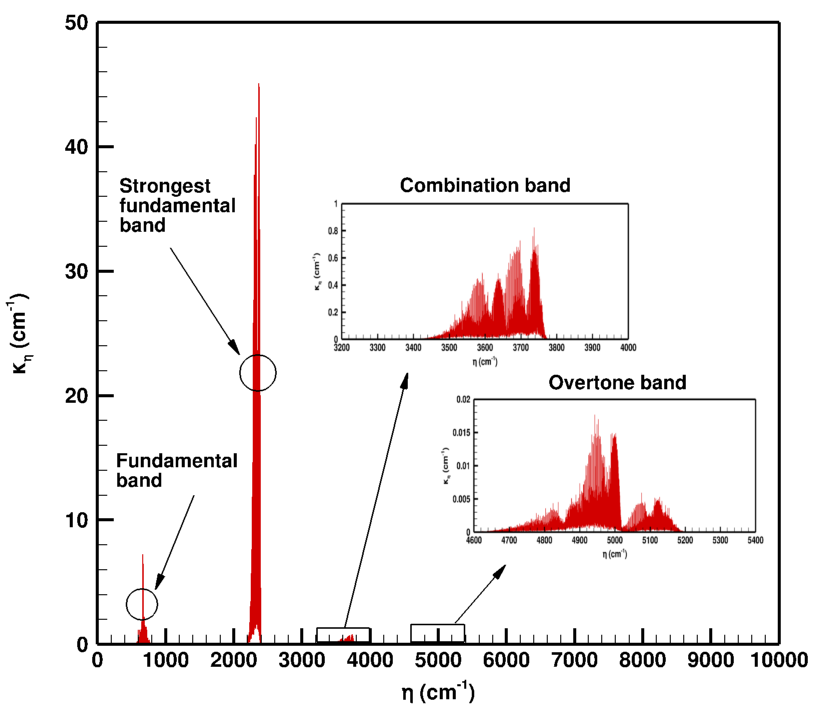}
    \caption{Spectral absorption coefficient of pure $CO_2$ at 1000 K and 1 atm}
\end{figure}

\begin{figure}[]
	\centering
  \begin{subfigure}[b]{0.4\textwidth}		
		\includegraphics[width=\textwidth]{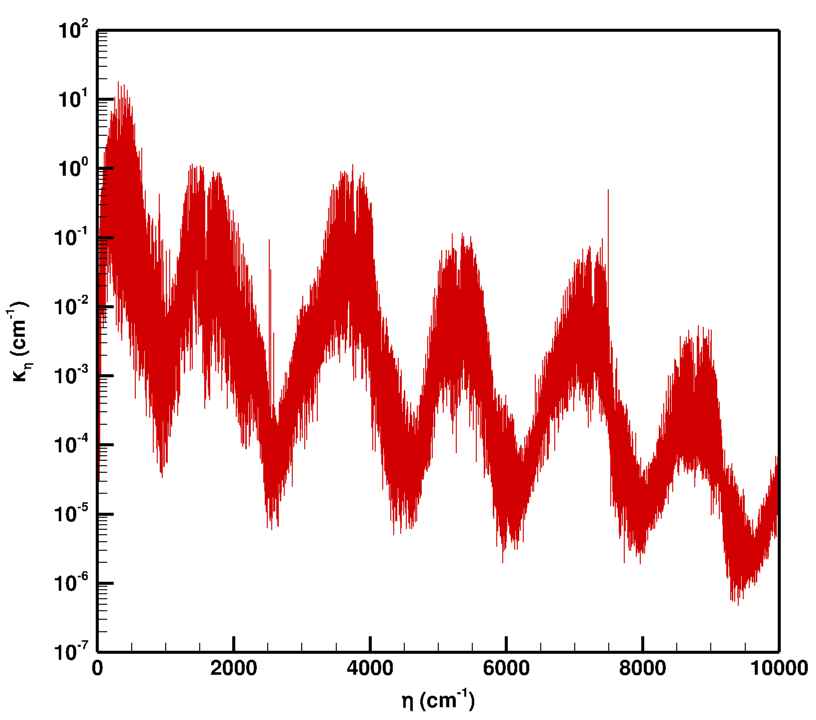}
		\caption{}
	\end{subfigure}
	\hspace{1.52cm}
  \begin{subfigure}[b]{0.4\textwidth}	
		\includegraphics[width=\textwidth]{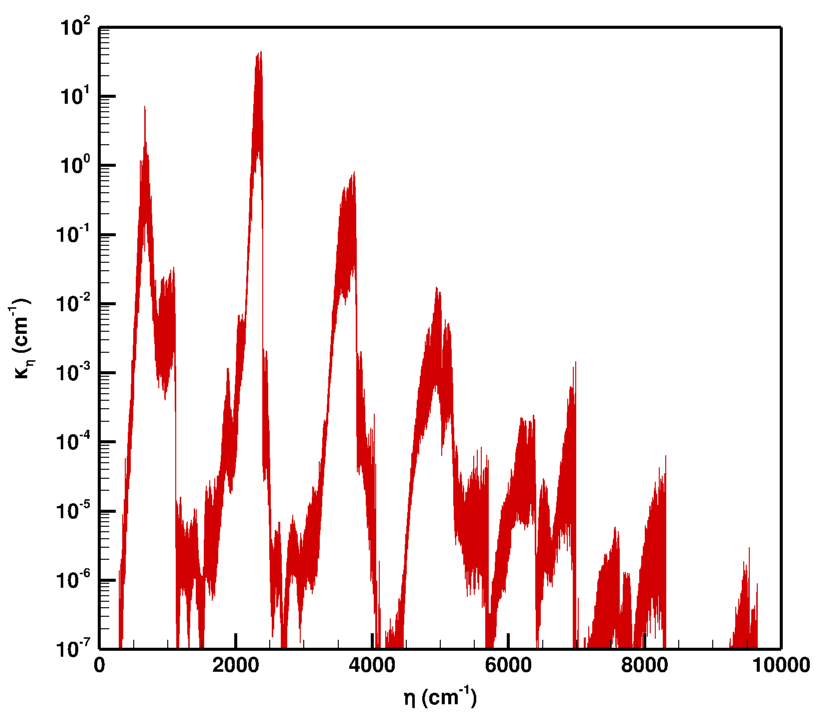}
		\caption{}
	\end{subfigure}
	\caption{Spectral absorption coefficient on $log$ scale at 1000 K and 1 bar for (a) pure $H_2O$ (b) pure $CO_2$}
\end{figure}

\subsection{FSK distribution of individual $H_2O$ and $CO_2$ gases}
The spectral absorption coefficient is re-ordered using FSK method into smooth $k$ vs $g$ distribution for pure $H_2O$ and $CO_2$ gases at 1000 K and 1 bar for single thermodynamic state of gas distributions. The $k$-distribution obtained by FSK method for both the gases is shown in Figure 4 and these distributions are monotonically increasing function. Similarly, the stretching function for $T_w=500$ and 1500, and $T_g=1000$ for $H_2O$ and $CO_2$ gas is shown in Figure 5, this distribution is quite erratic and $a$ should be selected carefully to obtain correct solution of RTE.

\begin{figure}[]
    \centering
	\begin{minipage}[t]{7cm} 
		\centering 
		\includegraphics[scale=0.23]{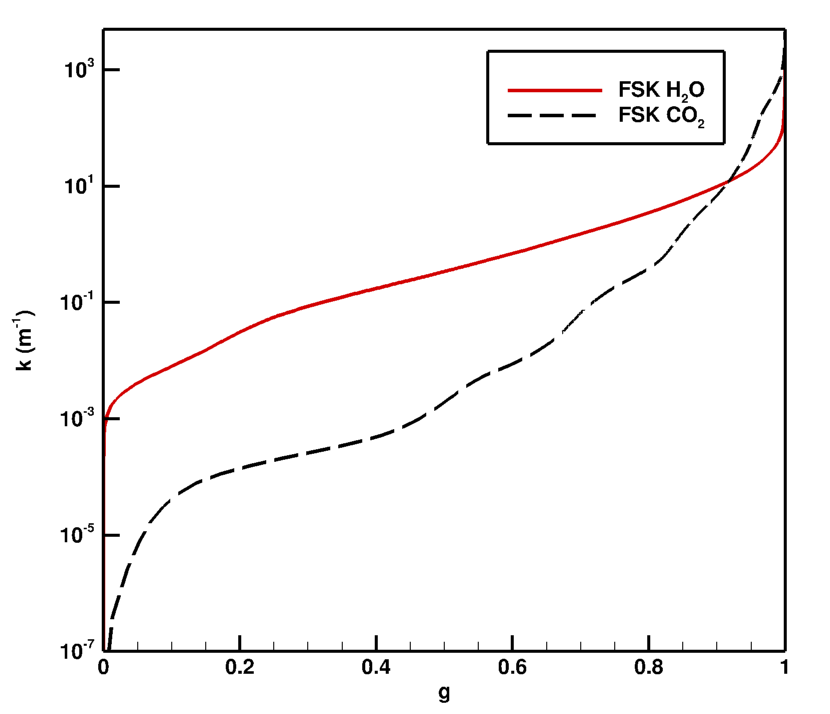} 
		\caption{$k$-distribution for pure $H_2O$ and $CO_2$ at 1000 K and 1 bar} 
	\end{minipage} 
	\hspace{2cm} 
	\begin{minipage}[t]{7cm} 
		\centering 
		\includegraphics[scale=0.23]{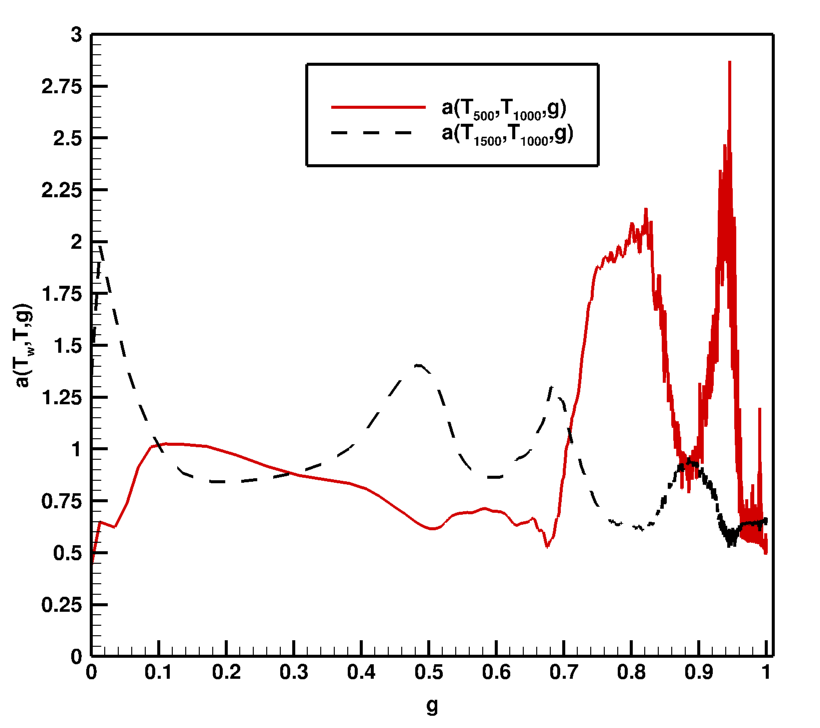} 
		\caption{Non-gray stretching factor $a(T_w,T,g)$} 
	\end{minipage}
	\end{figure}

\begin{table}[]
\centering
\caption{Band description of $H_2O$ and $CO_2$ gases}
\begin{tabular}{@{} l *3l @{}}
\toprule
\textbf{Species} & \textbf{Band}          & \textbf{Band Description} & \textbf{Wavenumber Range}                \\ \midrule
\multirow{3}{*}{$H_2O$} & 6.3 $\mu \:m$ & \begin{tabular}[l]{@{}l@{}}Strongest fundamental band\end{tabular}          & 1000 $cm^{-1}$ to 2100$cm^{-1}$ \\  
                        & 2.7 $\mu \:m$ & \begin{tabular}[l]{@{}l@{}}Strong overtone/fundamental band\end{tabular}    & 3000 $cm^{-1}$ to 4300$cm^{-1}$ \\  
        & 1.8 $\mu \:m$ & Combination band & 4800 $cm^{-1}$ to 5800$cm^{-1}$ \\ 
\multirow{4}{*}{$CO_2$} & 15 $\mu \:m$  & Fundamental band                                                              & 400 $cm^{-1}$ to 1200$cm^{-1}$  \\  
                        & 4.3 $\mu \:m$ & \begin{tabular}[l]{@{}l@{}}Strong fundamental/head-forming band\end{tabular} & 1900 $cm^{-1}$ to 2600$cm^{-1}$ \\ 
        & 2.7 $\mu \:m$ & Combination band & 3200 $cm^{-1}$ to 4000$cm^{-1}$ \\ 
        & 2.0 $\mu \:m$ & Overtone band    & 4600 $cm^{-1}$ to 5400$cm^{-1}$ \\ \bottomrule
\end{tabular}
\end{table}
\subsection{Solution of RTE for homogeneous isothermal and isobaric medium}
\subsubsection{Case 1}
The non-dimensional radiative heat flux at the bottom wall and the divergence of radiative heat flux along the horizontal line at mid height are calculated by solving RTE using FSK for different set of quadrature points for homogeneous isothermal and isobaric condition of a gas inside a square cavity. The cavity is filled with pure $CO_2$ at 1000 K and 1 bar and the walls of the cavity are black and cold. The results obtained using FSK are compared against LBL as shown in Figure 6 and 7. The non-dimensional flux and the divergence of radiative heat flux calculated with FSK almost matches with 16 and above quadrature points with the results of LBL method. There is a little inaccuracy for FSK with 12 points, which eventually vanishes by increasing number of quadrature points. Moreover, there is drastic reduction in computational time required to solve RTE, i.e., around 28000 times for FSK with maximum number of quadrature points ($p$=64) which is shown in Table 4. 

\begin{figure}[htbp]
    \centering
	\begin{minipage}[t]{7cm} 
		\centering 
		\includegraphics[scale=0.23]{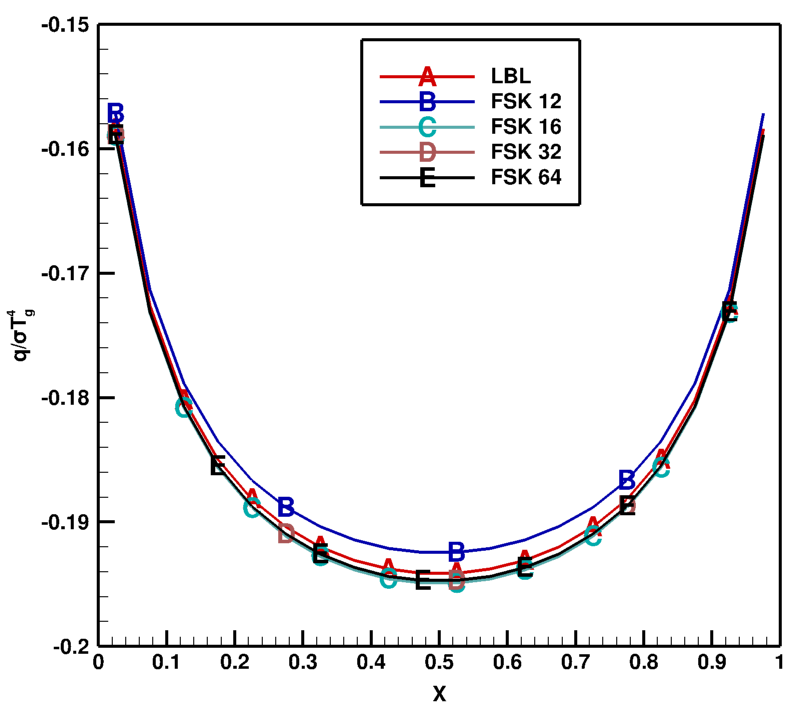} 
		\caption{Non-dimensional radiative heat flux on the bottom wall} 
	\end{minipage} 
	\hspace{2cm} 
	\begin{minipage}[t]{7cm} 
		\centering 
		\includegraphics[scale=0.23]{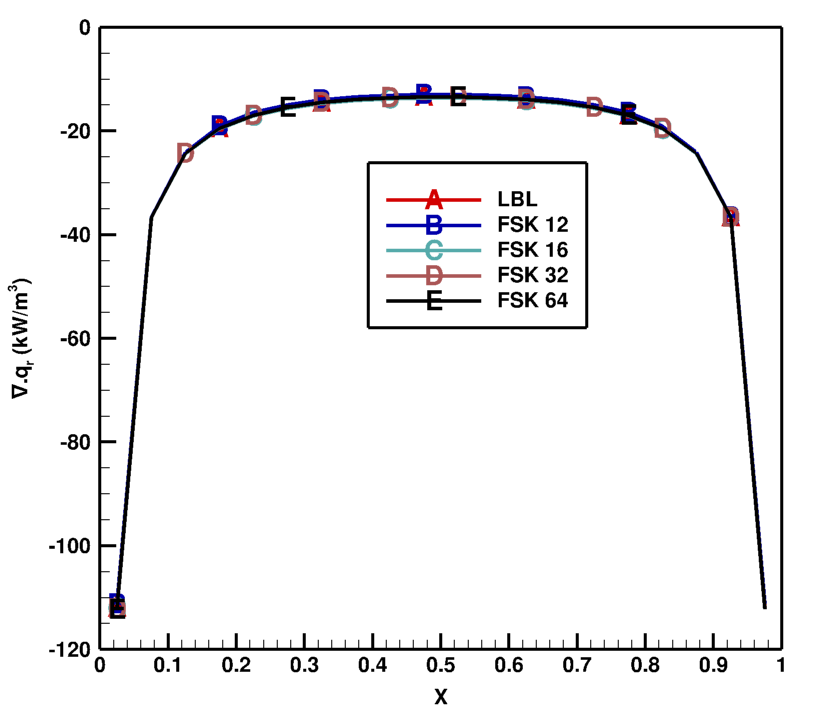} 
		\caption{Divergence of radiative heat flux along the horizontal line at the mid height of the cavity } 
	\end{minipage}
    
\end{figure}
\subsubsection{Case 2}
The above case is slightly modified by maintaining all the walls at 500 K and the rest conditions are kept same as in case 1. The stretching function is also required to calculate in this problem, as walls are at some temperature other than medium temperature. The non-dimensional radiative heat flux at the bottom wall and the divergence of radiative heat flux along the horizontal line at mid height is shown in Figures 8 and 9, respectively. The computational time required to solve RTE is shown in Table 5.
\subsubsection{Case 3}
Finally, the above case is made more realistic by keeping temperature at one wall as 1500K, other walls are maintained at 500 K and the rest conditions are kept same as that of case 1. The non-dimensional radiative heat flux at the bottom wall and the divergence of radiative heat flux along the horizontal line at mid height are shown in Figures 10 and 11, respectively. The computational time required to solve RTE is shown in Table 6. This case takes slightly more computational time compared with other two cases, this is due to calculation of stretching function for two different temperatures and solving RTE with two stretching functions.
\subsection{Solution of RTE for non-homogeneous non-isothermal non-isobaric medium}
The non-homogeneous non-isothermal and non-isobaric domain can be splitted into regions of homogeneous isothermal and isobaric medium and the properties of each region are taken from the look-up table for the available thermodynamic states and can be interpolated using multi-dimensional linear interpolation for unavailable thermodynamic states. Accuracy of RTE solution using multi-dimensional interpolation technique have been tested for three different test cases presented below.
\begin{table}[htbp]
\caption{Computational time required for the solution of RTE by different methods and quadrature points on 3.4 GHz i7 $4^{th}$ generation CPU}
\centering
\begin{tabular}{@{} l *1c @{}}
\toprule
\multicolumn{1}{l}{\textbf{Method}}    & \textbf{CPU Time (s)} \\
\midrule
LBL Calculations       & 5592                        \\ 
FSK with 12 $k$-points & 0.039                       \\ 
FSK with 16 $k$-points & 0.052                       \\ 
FSK with 32 $k$-points & 0.09                        \\ 
FSK with 64 $k$-points & 0.2                         \\ \bottomrule
\end{tabular}
\label{table:c1}
\end{table}
\newpage

\begin{figure}[]
    \centering
	\begin{minipage}[t]{7cm} 
		\centering 
		\includegraphics[scale=0.23]{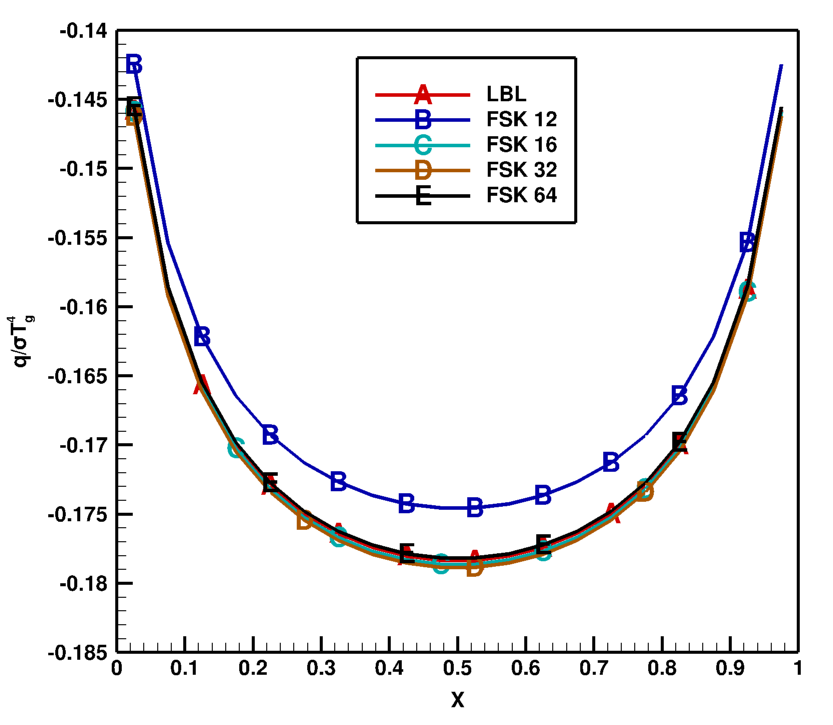} 
		\caption{Non-dimensional radiative heat flux on the bottom wall} 
	\end{minipage} 
	\hspace{2cm} 
	\begin{minipage}[t]{7cm} 
		\centering 
		\includegraphics[scale=0.23]{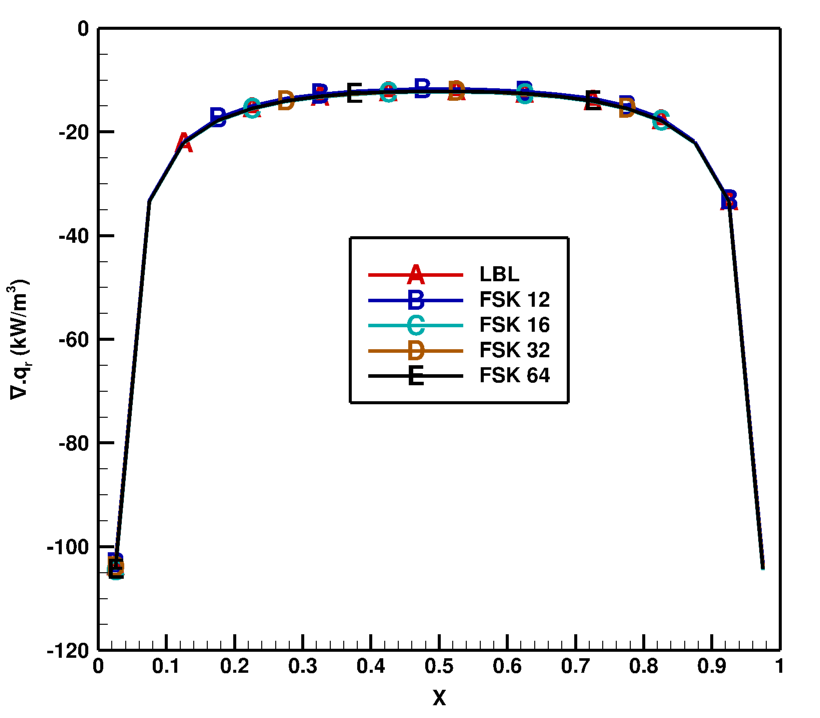} 
		\caption{Divergence of radiative heat flux along the horizontal line at the mid height of the cavity} 
	\end{minipage}
	\end{figure}

\begin{table}[]
\caption{Computational time required for the solution of RTE by different methods and quadrature points on 3.4 GHz i7 $4^{th}$ generation CPU}
\centering
\begin{tabular}{@{} l *1c @{}}
\toprule
\multicolumn{1}{l}{\textbf{Method}}    & \textbf{CPU Time (s)} \\
\midrule
LBL Calculations       & 5657                        \\ 
FSK with 12 $k$-points & 0.041                       \\ 
FSK with 16 $k$-points & 0.053                       \\ 
FSK with 32 $k$-points & 0.011                        \\ 
FSK with 64 $k$-points & 0.23                         \\ \bottomrule
\end{tabular}
\label{table:c2}
\end{table}

\begin{figure}[]
    \centering
	\begin{minipage}[t]{7cm} 
		\centering 
		\includegraphics[scale=0.23]{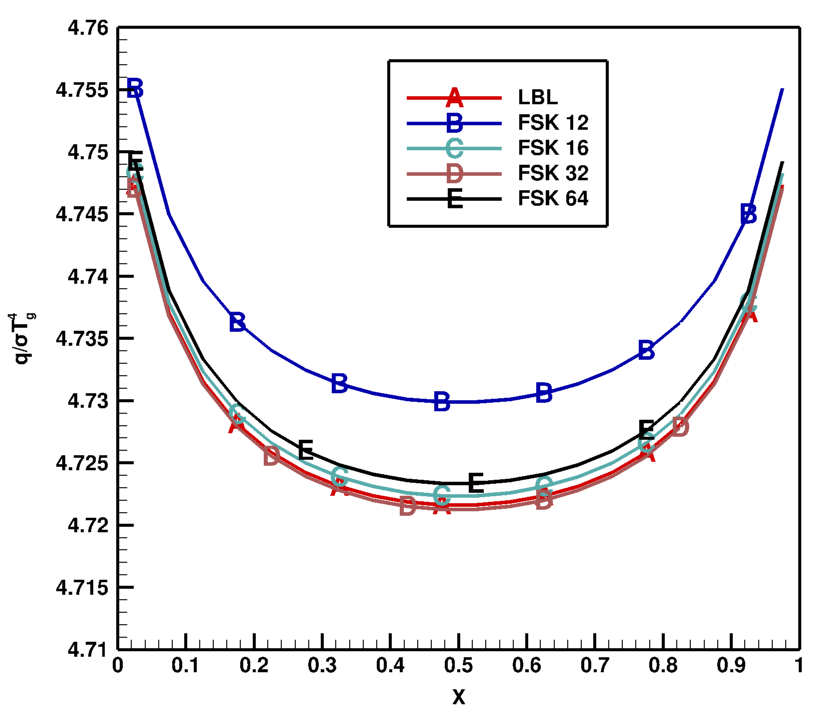} 
		\caption{Non-dimensional radiative heat flux on the bottom wall} 
	\end{minipage} 
	\hspace{2cm} 
	\begin{minipage}[t]{7cm} 
		\centering 
		\includegraphics[scale=0.23]{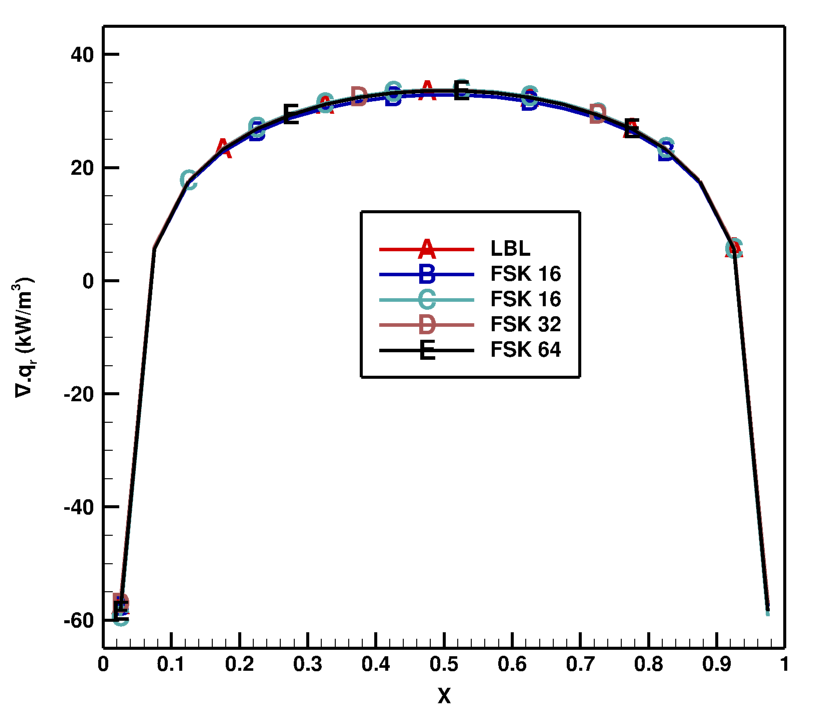} 
		\caption{Divergence of radiative heat flux along the horizontal line at the mid height of the cavity} 
	\end{minipage}
	\end{figure}

\begin{table}[]
\caption{Computational time required for the solution of RTE by different methods and quadrature points on 3.4 GHz i7 $4^{th}$ generation CPU}
\centering
\begin{tabular}{@{} l *1c @{}}
\toprule
\multicolumn{1}{l}{\textbf{Method}}    & \textbf{CPU Time (s)} \\
\midrule
LBL Calculations       & 5705                        \\ 
FSK with 12 $k$-points & 0.044                       \\ 
FSK with 16 $k$-points & 0.059                       \\ 
FSK with 32 $k$-points & 0.012                        \\ 
FSK with 64 $k$-points & 0.25                         \\ \bottomrule
\end{tabular}
\label{table:c3}
\end{table}

\begin{figure}[]
    \centering
    \includegraphics[scale=0.3]{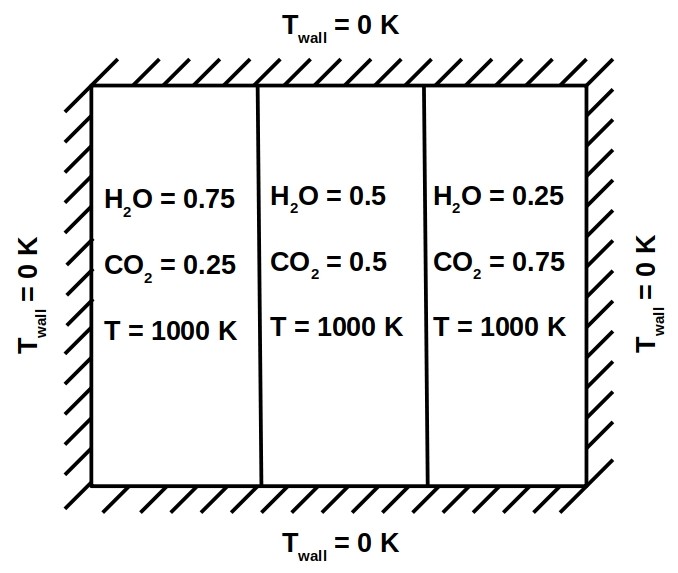}
    \caption{Geometry of trifurcated test case}
    \label{fig:tri}
\end{figure}
\begin{figure}[]
    \centering
	\begin{minipage}[t]{7cm} 
		\centering 
		\includegraphics[scale=0.23]{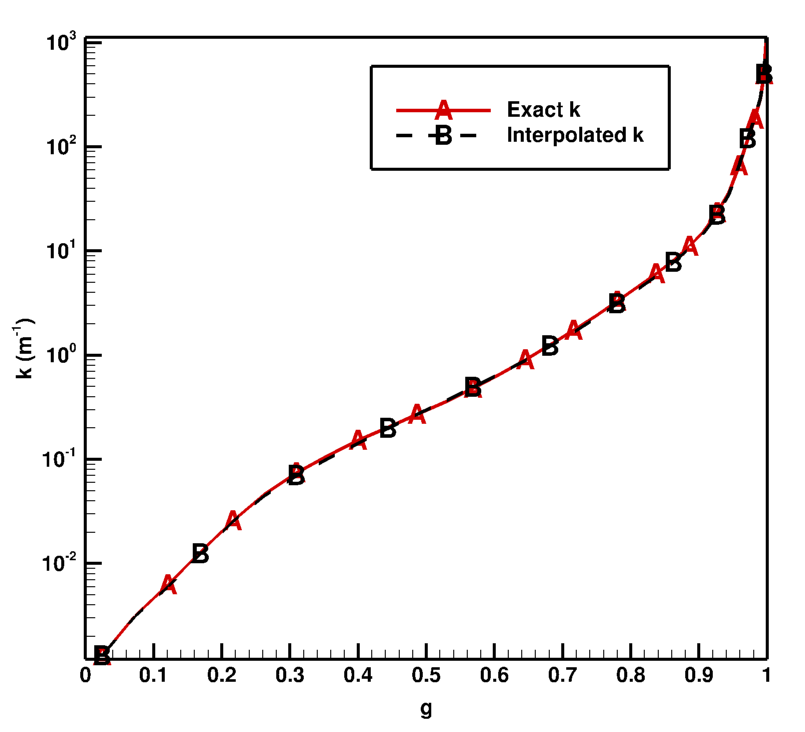} 
		\caption{Comparison of interpolated $k$-distribution obtained for middle section from FSK look-up table and exact FSK} 
	\end{minipage} 
	\hspace{2cm} 
	\begin{minipage}[t]{7cm} 
		\centering 
		\includegraphics[scale=0.23]{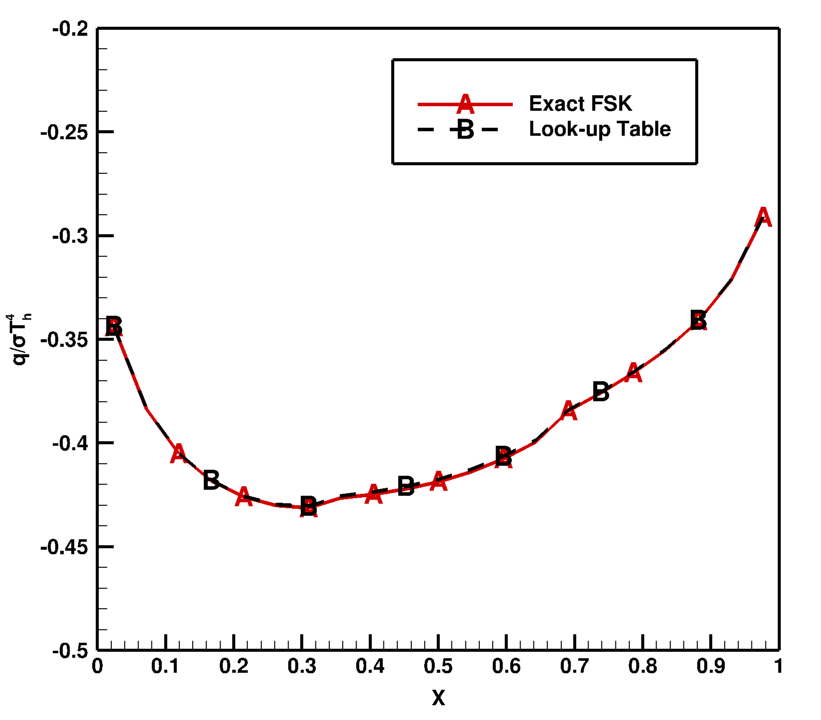} 
		\caption{Non-dimensional radiative heat flux at the bottom wall containing non-homogeneous isobaric and isothermal medium} 
	\end{minipage}
	\end{figure}

\subsubsection{Non-homogeneous isothermal and isobaric medium}
The computational domain used in this case contains $CO_2$ and $H_2O$ gases at different concentrations but at fixed temperature of 1000 K and pressure 1 bar. The non-homogeneous domain is trifurcated into the homogeneous regions as shown in Figure \ref{fig:tri}. The left side of the domain contains 25\% $CO_2$ and 75\% $H_2O$, middle part contains 50\% $CO_2$ and 50\% $H_2O$ and the right side contains 75\% $CO_2$ and 25\% $H_2O$. The walls of the cavity are black and cold. The radiative property of end regions is available in mixture FSK look-up table and the mid section is interpolated using a linear interpolation. The $k$-distribution and the non-dimensional radiative heat flux obtained from interpolation matches very well with exact FSK within 1\% error as shown in Figure 13 and 14, respectively. 
\subsubsection{Homogeneous isobaric and non-isothermal medium}
The computational domain used in this case contains 50\% $CO_2$ and 50\% $H_2O$ at 1 bar and at different temperatures. The non-isothermal domain is trifurcated into isothermal zones as in previous cases. The left, middle and the right regions of the domain are maintained at the temperature of 1000 K, 1100 K and 1200 K respectively. The walls of the cavity are black and cold. The property of the mid section is interpolated using a linear interpolation from the FSK look-up table. The $k$-distribution and the non-dimensional radiative heat flux obtained from interpolation matches very well with exact FSK as shown in Figure 15 and 16, respectively.

\begin{figure}[]
    \centering
	\begin{minipage}[t]{7cm} 
		\centering 
		\includegraphics[scale=0.23]{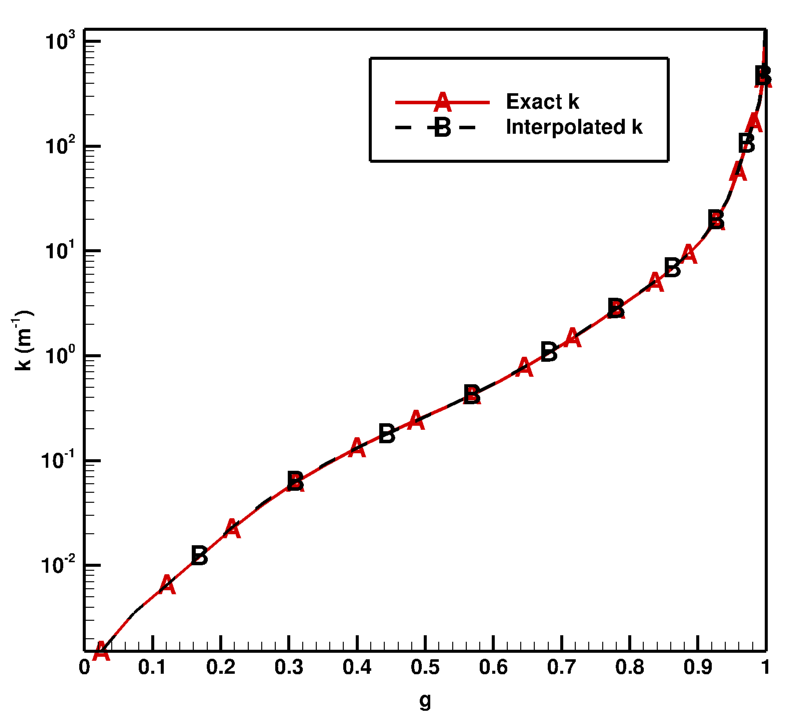} 
		\caption{Comparison of interpolated $k$-distribution obtained for middle section from FSK look-up table and exact FSK} 
	\end{minipage} 
	\hspace{2cm} 
	\begin{minipage}[t]{7cm} 
		\centering 
		\includegraphics[scale=0.23]{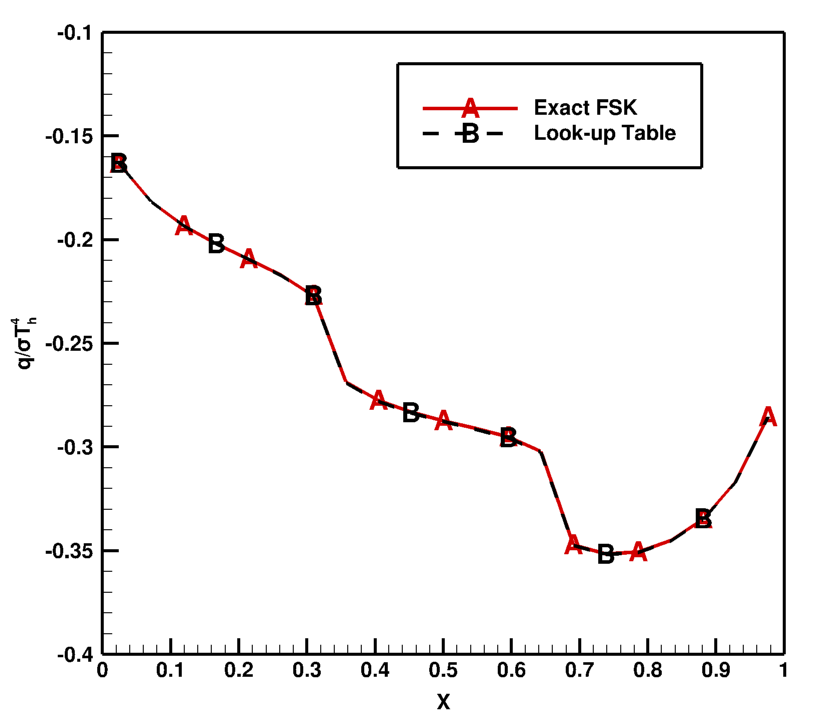} 
		\caption{Non-dimensional radiative heat flux at the bottom wall containing homogeneous isobaric and non-isothermal medium} 
	\end{minipage}
	\end{figure}

\subsubsection{Homogeneous isothermal and non-isobaric medium}
The computational domain used in this case contains 50\% $CO_2$ and 50\% $H_2O$ at 500K and at different pressures. The non-isobaric domain is trifurcated into isobaric zones as in previous cases in-order to test the accuracy of our look-up table in case of pressure variation. The left, middle and the right regions of the domain are maintained at the pressure of 5, 6 and 7 bar respectively. The walls of the cavity are black and cold. The property of the mid section is interpolated using a linear interpolation from the FSK look-up table. The $k$-distribution and the non-dimensional radiative heat flux obtained from interpolation matches very well with exact FSK as shown in Figure 17 and 18, respectively. The amount of radiative heat flux at the walls of the medium is quite high as can be seen from the magnitude of non-dimensional heat flux in Figure 18. This is due to the fact that, due to rise in pressure, medium becomes optically thick, which results into increase in the amount of radiative heat transfer.

\begin{figure}[]
    \centering
	\begin{minipage}[t]{7cm} 
		\centering 
		\includegraphics[scale=0.23]{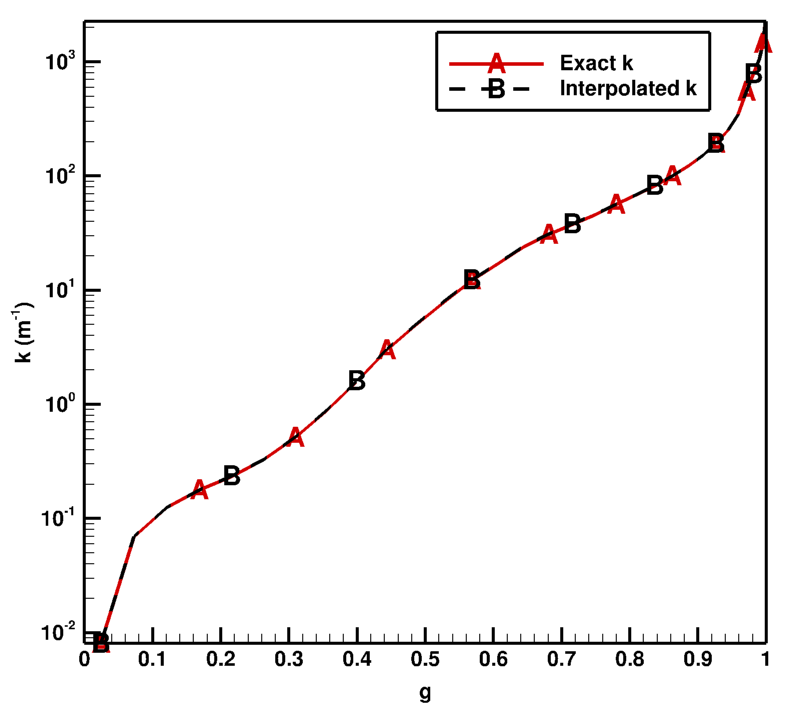} 
		\caption{Comparison of interpolated $k$-distribution obtained for middle section from FSK look-up table and exact FSK} 
	\end{minipage} 
	\hspace{2cm} 
	\begin{minipage}[t]{7cm} 
		\centering 
		\includegraphics[scale=0.23]{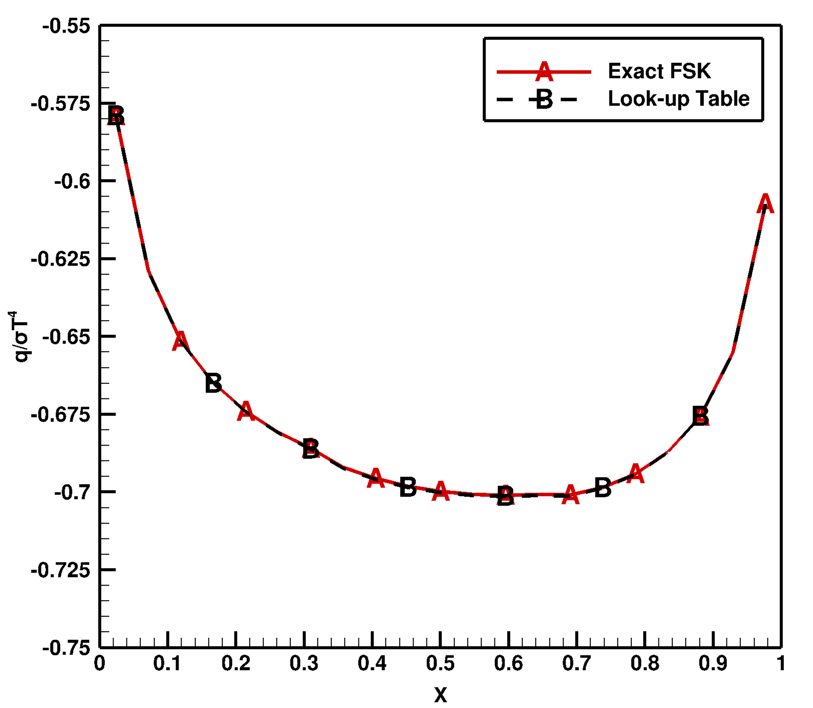} 
		\caption{Non-dimensional radiative heat flux at the bottom wall containing homogeneous isothermal and non-isobaric medium} 
	\end{minipage}
	\end{figure}

\subsubsection{Non-homogeneous isobaric and non-isothermal medium}
The computational domain used in this case contains $CO_2$ and $H_2O$ at different concentrations and temperatures. Again, the non-homogeneous and non-isothermal domain is trifurcated into homogeneous and isothermal zones as in previous cases. The left side of the domain contains 25\% $CO_2$ and 75\% $H_2O$ at 1000 K, middle part contains 50\% $CO_2$ and 50\% $H_2O$ at 1100 K and the right side contains 75\% $CO_2$ and 25\% $H_2O$ at 1200 K and at a pressure of 1 bar. The walls of the cavity are black and cold. The property of the mid section is interpolated using a bi-linear interpolation from the FSK look-up table. The $k$-distribution and the non-dimensional radiative heat flux obtained from interpolation matches very well with exact FSK as shown in Figure 19 and 20, respectively.

\subsection{Solution of RTE for mixture of gases}
The solution of RTE for the mixture of gases in the previous section have been obtained by constructing the $k$-distribution for mixture spectral absorption coefficients. However, this is not always possible, here, we are evaluating the performance of  three different mixing models namely, SMM, MMM and HMM which use the $k$-distribution of individual gas, in the following sections.

\begin{figure}[]
    \centering
	\begin{minipage}[t]{7cm} 
		\centering 
		\includegraphics[scale=0.23]{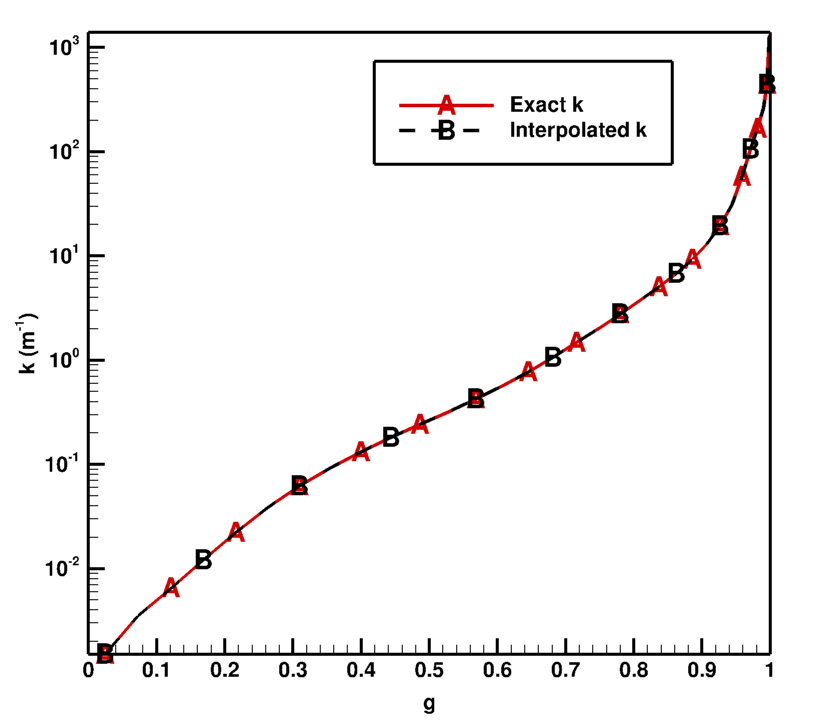} 
		\caption{Comparison of interpolated $k$-distribution obtained for middle section from FSK look-up table and exact FSK} 
	\end{minipage} 
	\hspace{2cm} 
	\begin{minipage}[t]{7cm} 
		\centering 
		\includegraphics[scale=0.23]{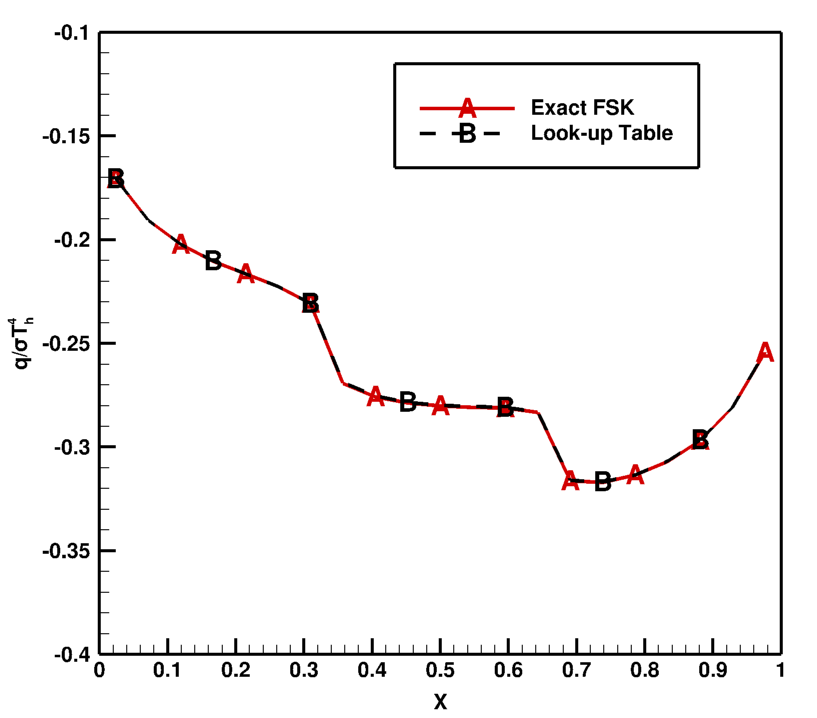} 
		\caption{Non-dimensional radiative heat flux at the bottom wall containing non-homogeneous and non-isothermal medium} 
	\end{minipage}
	\end{figure}

\subsubsection{Homogeneous isobaric and isothermal medium}
A square cavity whose walls are black and cold contains contains mixture of 50\% $H_2O$ and 50\% $CO_2$ at 1000 K and 1 atm pressure. Figure 21 shows spectral absorption coefficient of individual $CO_2$ and $H_2O$ gases where significant overlapping on the spectral scale is seen and corresponding FSK obtained from different mixing models is shown in Figure 22. The $k$-distribution assembled by spectral addition method (SAM) is the most accurate technique and used for benchmarking of other models. It can be seen that superposition mixing model (SMM) deviates a lot from SAM at lower $g$-values because of significant overlap of spectral lines at lower absorption coefficient values. At higher $g$-values, i.e., $g>$0.9, SMM overlaps with SAM. For multiplication mixing model (MMM) the trend is reverse when compared to SMM. The $k$-distribution profile by hybrid mixing model (HMM) is between FSK-SMM and FSK-SMM. 

The non-dimensional radiative heat flux at the bottom wall and the divergence of radiative heat flux along the horizontal line at mid height is shown in Figure 23 and 24, respectively.
The FSK with SAM is found to be most accurate when compared to LBL method. FSK-MMM gives good results when compared with FSK-SAM with maximum error of 0.6\%, FSK-SMM is least accurate with maximum error of 12\% and FSK-HMM lies in between FSK-SMM and FSK-MMM with the maximum error of 6\%.

\begin{figure}[]
    \centering
	\begin{minipage}[t]{7cm} 
		\centering 
		\includegraphics[scale=0.23]{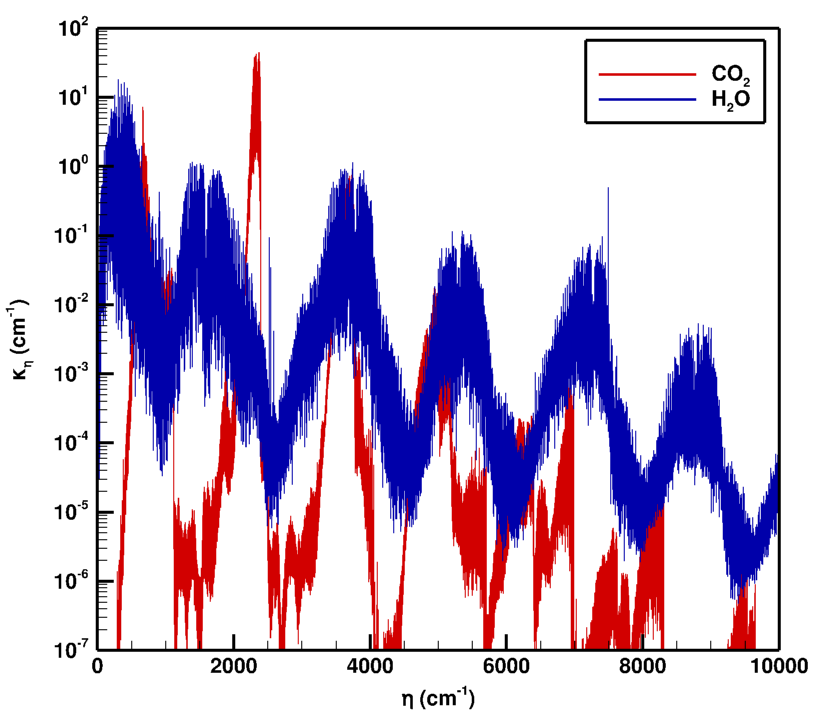} 
		\caption{Spectral absorption coefficient of individual $CO_2$ and $H_2O$ gases at 1000 K and 1 bar} 
	\end{minipage} 
	\hspace{2cm} 
	\begin{minipage}[t]{7cm} 
		\centering 
		\includegraphics[scale=0.23]{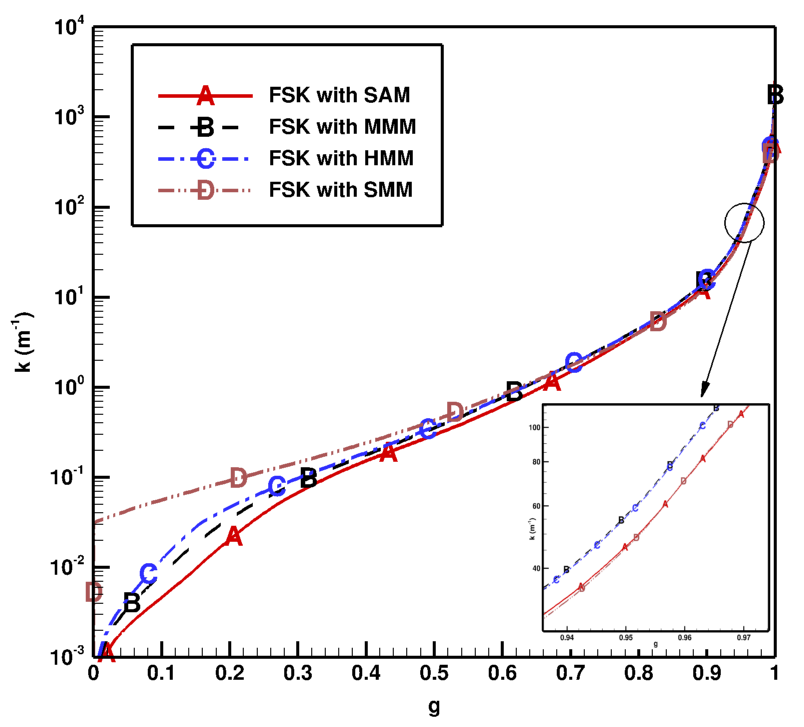} 
		\caption{Comparison of $k$-distribution obtained from different gas mixing models} 
	\end{minipage}
	\end{figure}

\begin{figure}[]
    \centering
	\begin{minipage}[t]{7cm} 
		\centering 
		\includegraphics[scale=0.23]{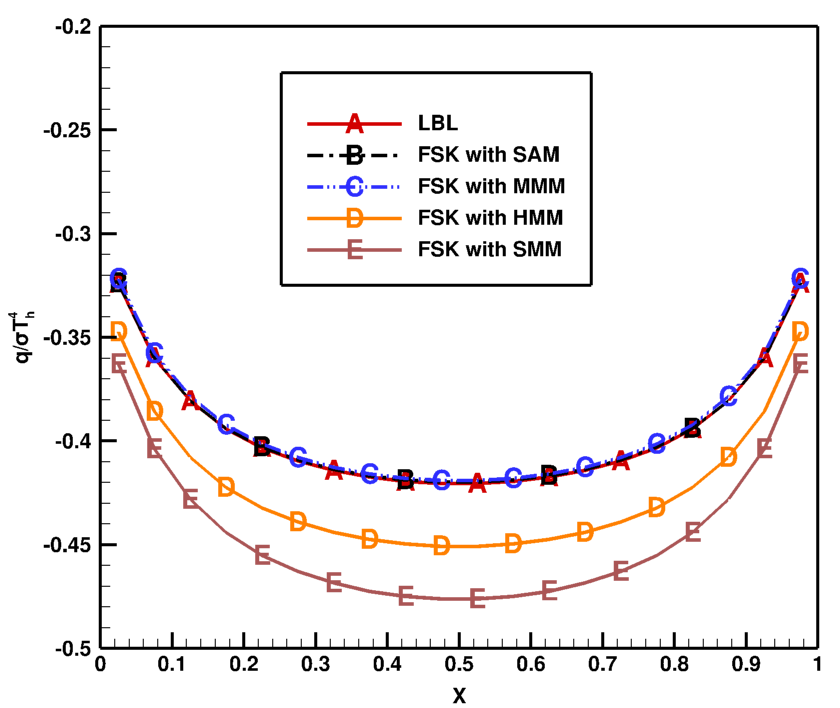} 
		\caption{Non-dimensional radiative heat flux on the bottom wall} 
	\end{minipage} 
	\hspace{2cm} 
	\begin{minipage}[t]{7cm} 
		\centering 
		\includegraphics[scale=0.23]{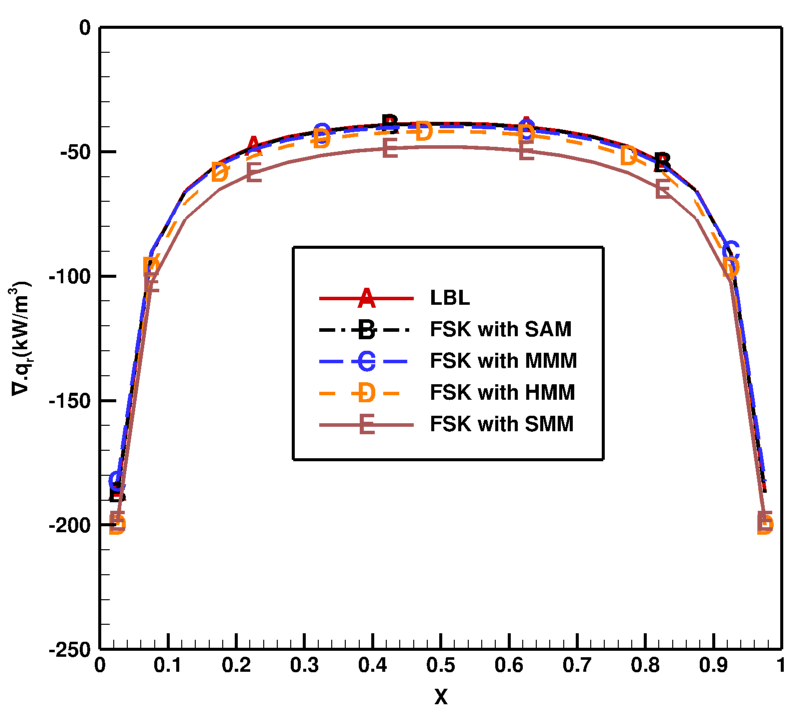} 
		\caption{Divergence of radiative heat flux along the horizontal line at the mid height of the cavity} 
	\end{minipage}
	\end{figure}

\subsubsection{Non-homogeneous isobaric and non-isothermal medium}
A square cavity has been bifurcated into two regions whose left part contains 25\% $CO_2$ and 75\% $H_2O$ at 1000 K and right part contains 75\% $CO_2$ and 25\% $H_2O$ at 1200 K and 1 atm pressure. The walls of the cavity are cold and black. Figure 25 and 26 show the non-dimensional radiative heat flux at the bottom wall and the divergence of radiative heat flux along the horizontal line at the mid height of the cavity. Among all the mixing models, MMM gives good accuracy in comparison to SAM followed by HMM and SMM with the maximum error 2\%, 2.5\% and 4\%, respectively.

\begin{figure}[]
    \centering
	\begin{minipage}[t]{7cm} 
		\centering 
		\includegraphics[scale=0.23]{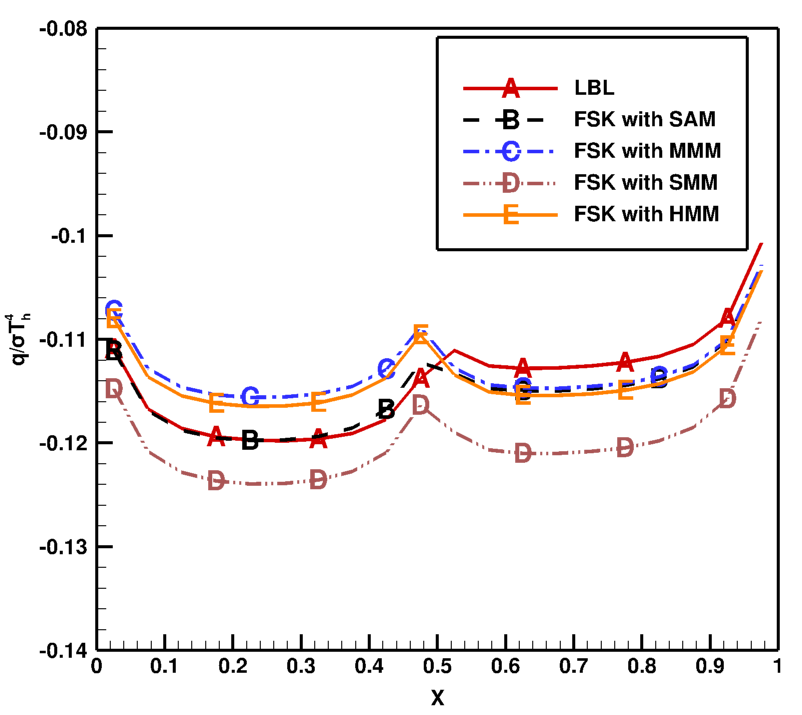} 
		\caption{Non-dimensional radiative heat flux on the bottom wall} 
	\end{minipage} 
	\hspace{2cm} 
	\begin{minipage}[t]{7cm} 
		\centering 
		\includegraphics[scale=0.23]{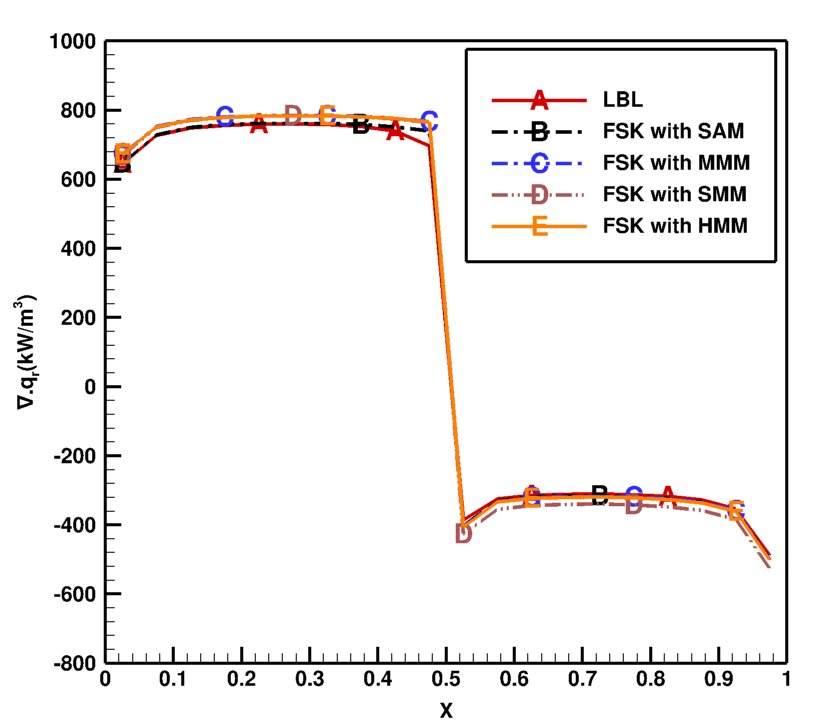} 
		\caption{Divergence of radiative heat flux along the horizontal line at the mid height of the cavity} 
	\end{minipage}
	\end{figure}

\section{Conclusions}

The non-gray radiative properties of participating gases namely, $CO_2$ and $H_2O$ have been calculated from the HITEMP-2010 database and line-by-line (LBL) calculation of radiative transfer equation has been performed to obtain the radiative heat flux at the bottom wall of the cavity and the divergence of radiative flux at the horizontal line at the mid height of the cavity. This solution has been used for the benchmarking the methods of full spectrum $k$-distribution method (FSK) for homogeneous isothermal isobaric medium and non-homogeneous non-isothermal non-isobaric medium, and also for the single gas and mixture of gases. The FSK method is almost exact and highly efficient for single gas and homogeneous isothermal isobaric medium. It drastically reduces the computational time required to solve RTE i.e., around 0.6 million times in comparison to LBL method. This method is further extended to non-homogeneous non-isobaric and non-isothermal medium by developing the FSK look-up table for radiative properties. An efficient multidimensional linear interpolation method is also proposed for non-available data in the FSK look-up table. Furthermore, some gas mixing methods namely SMM, MMM, HMM have been explored for the calculation of radiative heat transfer for the mixture of gases from the $k$-distribution of single gas and the radiative transfer calculation have been performed in a cavity. The accuracy of these mixing models are almost same in comparison to FSK for mixture of gases. These models have reduced the computational resource requirements drastically with almost same accuracy of LBL method, thus can be used with other models of fluid flow and heat transfer for the applications of combustion, plume radiation and gasifier etc.
\bibliography{mybibfile}
\end{document}